\documentclass{nature}
\usepackage{graphicx}
\usepackage{amsmath}
\usepackage{color}
\usepackage{physics}
\usepackage{hyperref}
\usepackage{color}
\usepackage{soul}
\usepackage{slashed}
\usepackage{amssymb}
\usepackage[labelformat=empty, hypcap=false]{caption}
\newcommand{\red}[1]{{\color{black} #1}}

\usepackage{amsfonts}
\usepackage[version=3]{mhchem}

\makeatletter
\let\saved@includegraphics\includegraphics

\title{Honeycomb Layered Oxides With Silver Atom Bilayers and Emergence of Non-Abelian SU(2) Interactions}

\author{Titus Masese$^{1,2}$, Godwill Mbiti Kanyolo$^3$, Yoshinobu Miyazaki$^4$, Miyu Ito$^4$, Noboru Taguchi$^1$, Josef Rizell$^{1,5}$, Shintaro Tachibana$^6$, Kohei Tada$^1$, Zhen-Dong Huang$^7$, Abbas Alshehabi$^8$, Hiroki Ubukata$^9$, Keigo Kubota$^1$, Kazuki Yoshii$^1$, Hiroshi Senoh$^1$, Cédric Tassel$^9$, Yuki Orikasa$^6$, Hiroshi Kageyama$^9$ \& Tomohiro Saito$^4$}

\begin{document}

\maketitle

\begin{affiliations}

\item Research Institute of Electrochemical Energy, National Institute of Advanced Industrial Science and Technology (AIST), 1-8-31 Midorigaoka, Ikeda, Osaka 563-8577, JAPAN

\item AIST-Kyoto University Chemical Energy Materials Open Innovation Laboratory (ChEM-OIL), Sakyo-ku, Kyoto 606-8501, JAPAN

\item Department of Engineering Science, The University of Electro-Communications, 1-5-1 Chofugaoka, Chofu, Tokyo 182-8585, JAPAN

\item Tsukuba Laboratory, Technical Solution Headquarters, Sumika Chemical Analysis Service (SCAS), Ltd.,Tsukuba, Ibaraki 300-3266, JAPAN 

\item Department of Physics, Chalmers University of Technology, SE-412 96 G\"{o}teborg, SWEDEN

\item Graduate School of Life Sciences, Ritsumeikan University, 1-1-1 Noji-higashi, Kusatsu, Shiga 525-8577, JAPAN

\item Key Laboratory for Organic Electronics and Information Displays and Institute of Advanced Materials (IAM), Nanjing University of Posts and Telecommunications (NUPT), Nanjing, 210023, CHINA

\item Department of Industrial Engineering, National Institute of Technology (KOSEN), Ibaraki College, 866 Nakane, Hitachinaka, Ibaraki 312-8508 JAPAN

\item Department of Energy and Hydrocarbon Chemistry, Graduate School of Engineering, Kyoto University, Nishikyo–ku, Kyoto 615–8510, JAPAN

\end{affiliations}

\begin{abstract}
Honeycomb layered oxides with monovalent or divalent, monolayered cationic lattices generally exhibit myriad crystalline features encompassing rich electrochemistry, geometries and disorders, which particularly places them as attractive material candidates for next-generation energy storage applications. Herein, we report global honeycomb layered oxide compositions, ${\rm Ag_2}M_2{\rm TeO_6}$ ($M = \rm Ni, Mg, \textit{etc}.$) exhibiting $\rm Ag$ atom bilayers with sub-valent states within Ag-rich crystalline domains of ${\rm Ag_6}M_2{\rm TeO_6}$ and $\rm Ag$-deficient domains of ${\rm Ag}_{2 - x}\rm Ni_2TeO_6$ ($0 < x < 2$). The $\rm Ag$-rich material characterised by aberration-corrected transmission electron microscopy reveals local atomic structural disorders characterised by aperiodic stacking and incoherency in the bilayer arrangement of $\rm Ag$ atoms. Meanwhile, the global material not only displays high ionic conductivity, but also manifests oxygen-hole electrochemistry during silver-ion extraction. Within the $\rm Ag$-rich domains, the bilayered structure, argentophilic interactions therein and the expected $\rm Ag$ sub-valent states ($1/2+, 2/3+,$ \textit{etc}.) are theoretically understood via spontaneous symmetry breaking of SU($2$)$\times$U($1$) gauge symmetry interactions amongst $3$ degenerate mass-less chiral fermion states, justified by electron occupancy of silver $4d_{z^2}$ and $5s$ orbitals on a bifurcated honeycomb lattice. This implies that bilayered frameworks have research applications that go beyond the confines of energy storage. 

\end{abstract}
\begin{figure*}
\centering
\includegraphics[width=\textwidth,clip=true]{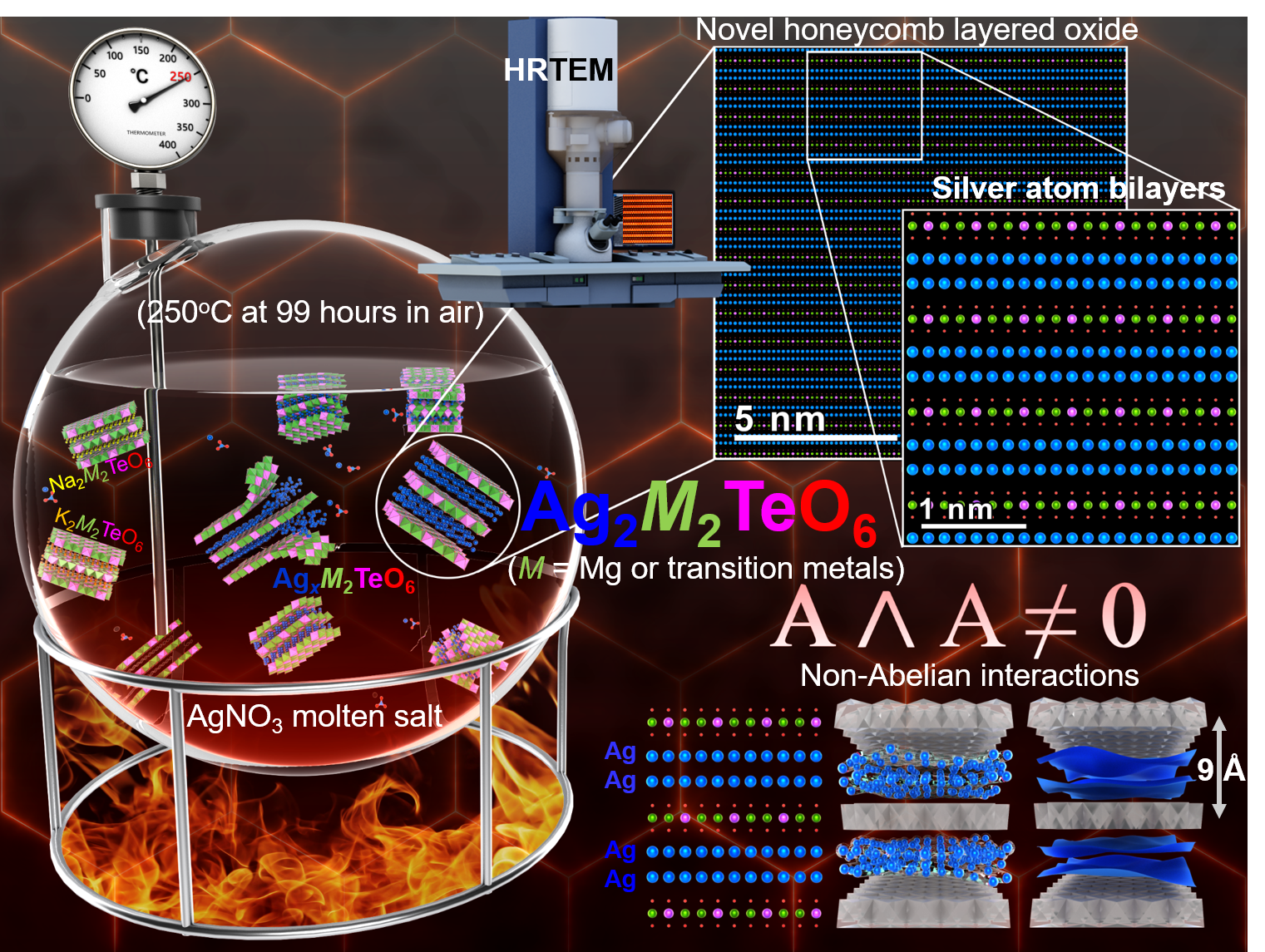}
\end{figure*}

\newpage

\section{\label{section: Introduction} Introduction}

Advancements in nanotechnology have unearthed a trove of multifunctional materials that promise to redefine the frontiers of research and applications with quixotic-like physical, electrochemical, and structural functionalities.\cite{kanyolo2021honeycomb, kanyolo2022advances} Recent exemplars of these capabilities are embodied by honeycomb layered oxides, which exhibit unique electronic and magnetic behaviour, fast ion kinetics, exotic geometries and phase transitions\cite{kanyolo2021honeycomb, kanyolo2022advances}, alongside desirable electrochemical properties for energy storage applications.\cite{house2020, maitra2018, wang2018a, yabuuchi2014, cabana2013, song2019} In particular, these materials mainly feature a monolayer of monovalent or divalent atoms (such as $\rm Li, Na, K, Ag$ and $\rm Cu$), typically in a hexagonal (triangular) or honeycomb lattice, sandwiched between hexagonal or honeycomb transition metal - or heavy metal oxides, rendering them ideal for the design of next-generation multifunctional materials. The honeycomb and/or hexagonal lattice is visible in the various two-dimensional (2D) slices of the material since these materials, despite having different types of atoms, are constituted solely by the face-centred cubic (FCC) packing and/or hexagonal close packing (HCP), which have been mathematically proven (assuming congruent spheres) to be the only optimised lattices in the three-dimensional (3D) sphere packing problem.\cite{hales2011revision} As a consequence, every 2D slice is either a hexagonal or honeycomb lattice, regardless of the type of atom constituting the slice one selects, unless there are deviations from the optimal condition, characterised by disclinations, distortions, dislocations, vacancies and/or other topological features.\cite{kanyolo2022conformal} Moreover, their unique crystalline structure and inherent structural symmetries facilitate 2D atomistic interactions to dominate the honeycomb layered heterostructures, which fosters the exploration not only of unconventional magnetic phenomena such as Heisenberg-Kitaev interactions\cite{kitaev2006anyons, kanyolo2021honeycomb, kanyolo2022advances}, but also new-fangled emergent properties such as quantum geometries and topologies.\cite{kanyolo2020idealised, kanyolo2022cationic, masese2021mixed, masese2021topological, kanyolo2021partition} Indeed, honeycomb layered tellurates (particularly, $\rm {\it A}_2{\it M}_2TeO_6$ (where $A = \rm Li, Na, K$ and $M = \rm Ni, Co, Mg, \textit{\textit{etc}.}$) compositions) inferred experimentally\cite{grundish2019electrochemical, kumar2013formation, evstigneeva2011new, sankar2014crystal, berthelot2012studies, masese2018rechargeable, masese2019high, yoshii2019sulfonylamide} and from computations\cite{tada2022implications} proffer a promising odyssey of probing into the functionalities of unchartered compositions that not only accommodate the aforementioned monolayer arrangement of monovalent or divalent atoms, but also the possibility of multilayered structures of sub-valent coinage metal atoms.\cite{johannes2007formation, yoshida2020static, taniguchi2020butterfly, yoshida2011novel, matsuda2012partially, yoshida2008unique, yoshida2006spin} 

Thus, the prospect of expounding the compositional diversity for honeycomb layered tellurates hosting coinage metal atoms (such as $\rm Ag, Cu$ and $\rm Au$) is poised to unlock new applications for this class of materials. Most notably, honeycomb layered tellurates that can accommodate a monolayer arrangement of $\rm Ag$ atoms have been envisioned to form structural coordinations that are very distinct from the typical prismatic and octahedral coordinations observed in alkali atoms (A compendium of the various slab arrangements (stackings) observed in honeycomb layered oxides is provided in the \textbf{Supplementary Information} section (\textbf {Supplementary Figure 1})). In this vein, $\rm Ag$ atoms in $\rm Ag$-based honeycomb layered oxides such as $\rm Ag_3Ni_2SbO_6$, $\rm Ag_3Ni_2BiO_6$, amongst others have been reported to form 
dumbbell/linear coordinations with two oxygen atoms in varied stacking arrangements.\cite{kanyolo2021honeycomb, kanyolo2022advances, zvereva2016orbitally, berthelot2012new, bette2019crystal, bhardwaj2014evidence} However, our interest in $\rm Ag$-based honeycomb layered tellurates was piqued by their propensity to adopt other variegated coordinations 
such as prismatic coordination, as predicted by theoretical studies.\cite{tada2022implications} In particular, the possible formation of an assortment of $\rm Ag$-atom structures can be traced to their anomalous valency states (\textit{i.e.}, valency states of between $0$ and $1+$ (technically referred to as sub-valent)), which have been posited to precipitate idiosyncratic structural and bonding properties when sandwiched between transition metal layers, such as the bilayer $\rm Ag$ atom arrangement observed in layered oxides such as in ${\rm Ag_2}M\rm O_2$ ($M = \rm Cr, Co, Ni, \textit{etc}.$), amongst others.\cite{johannes2007formation, taniguchi2020butterfly, yoshida2020static, yoshida2011novel, matsuda2012partially, yoshida2008unique, yoshida2006spin, beesk1981x, ahlert2003ag13oso6, kovalevskiy2020uncommon, kohler1985electrical}

In a bid to gain insights into the peculiar structural dispositions of $\rm Ag$-atoms, we report for the first time the synthesis and structural characterisation of honeycomb layered tellurates with global compositions, ${\rm Ag_2}M_2{\rm TeO_6}$ ($M = \rm Ni, Mg$, and other transition metal atoms) exhibiting $\rm Ag$ atom bilayers within $\rm Ag$-rich crystalline domains of ${\rm Ag_6}M_2{\rm TeO_6}$. Through aberration-corrected scanning transmission electron microscopy, we report and elucidate the intricate atomic disordered structure of ${\rm Ag_6}M_2{\rm TeO_6}$, which is noted to predominantly comprise triangular $\rm Ag$-atom bilayer lattices sandwiched between transition metal slabs with an aperiodic stacking sequence. Electrochemical measurements reveal both $\rm Ag_2Ni_2TeO_6$ and $\rm Ag_2Mg_2TeO_6$ to display $\rm Ag$-ion extraction electrochemistry marked by a predominant formation of oxygen holes that debilitates reversible $\rm Ag$-ion electrochemistry. Nonetheless, these global compositions exhibit relatively high ionic conductivities of $2.39 \times 10^{-2} \,\,\rm S\,cm^{-1}$ and $3.84 \times 10^{-4} \,\,\rm S\,cm^{-1}$ respectively at 100 $^{\circ}$C — comparable to those of canonical $\rm Ag$ superionic conductors reported to date.\cite{hull2000structural, hull2001structural, hull2002crystal, hull2002structural, nilges2004structure, nilges2005structures, matsunaga2004structural, hull2005ag+, lange2006ag10te4br3, lange2007polymorphism, angenault1989conductivite, rao2005preparation, daidouh2002structural, daidouh1997structure, fukuoka2003crystal, quarez2009crystal} 

Finally, the bilayered structure observed in the Ag-rich crystalline domains is theoretically understood by considering 3 degenerate mass-less chiral fermion states of silver given by left-handed states labelled by $\rm Ag_{+1/2}$ ($4d^{10}s^1$) and $\rm Ag_{-1/2}$ ($4d^95s^2$) treated as emergent iso-spin up ($+1/2$) and down ($-1/2$) degrees of freedom characteristic of special unitary group of degree 2 (SU($2$)) gauge symmetry (responsible for $\rm Ag$ oxidation states $\rm Ag^{1+}$ ($4d^{10}5s^0$) and $\rm Ag^{1-}$ ($4d^{10}5s^2$) respectively) and a right-handed oxidation state, $\rm Ag_0 \rightarrow Ag^{2+}$ ($4d^95s^2 \rightarrow 4d^95s^0$) on the honeycomb lattice, based on the occupancy of their $4d_{z^2}$ and $5s$ orbitals. Note that, $\rm Ag_{-1/2}$ and $\rm Ag_0$ are degenerate with essentially the same electronic state, $4d^95s^2$, \textit{albeit} form different oxidation states. Moreover, the oxidation states in the superscript also correspond to their respective valency states, achieved by a broken local SU($2$)$\times$U($1$) gauge symmetry.\cite{zee2010quantum} Here, unitary group of degree 1 (U($1$)) signifies the electromagnetic/Maxwell theory responsible for the electric charges of all the degenerate Ag states, whereas SU($2$) is the emergent gauge group of the additional interaction between the left-handed degenerate states, characteristic of $sd$-hybridisation, and analogous to lepton interactions in electroweak theory.\cite{zee2010quantum, weinberg1967model} Breaking this symmetry introduces effective (sub-valent) states such as $1/2+$ and $2/3+$ and a mass term between $\rm Ag^{2+}/Ag_0$ and $\rm Ag^{1-}/Ag_{-1/2}$, computed as the $\rm Ag-Ag'$ argentophilic interaction responsible for stabilising the observed bilayered structure, leaving $\rm Ag^{1+}/Ag_{+1/2}$ mass-less. Other considerations such as modular and conformal symmetry\cite{kanyolo2022cationic, kanyolo2022conformal} shed light on the nature of the bilayer, also observed in the other frameworks such as $\rm Ag_2^{1/2+}F^{1-}$, $\rm Ag_2^{1/2+}Ni^{3+}O_2^{2-}$ and the hybrid, $\rm Ag_3^{2/3+}Ni_2^{3+}O_4^{2-}$.\cite{schreyer2002synthesis, yoshida2006spin, johannes2007formation, sorgel2007ag3ni2o4} Ultimately, we regard the silver-based honeycomb layered tellurate as a pedagogical platform for further inquiry into the role of geometric features and non-commutative electromagnetic interactions, which go beyond energy storage applications.\cite{kanyolo2020idealised, kanyolo2022cationic}

\red{Throughout the paper and the \textbf{Supplementary Information}, we have adopted the notation: ${\rm Ag}_x{\rm Ni_2TeO_6}$ where the global material composition (as ascertained by inductively-coupled plasma atomic emission spectroscopy (ICP-AES)) is given by $x = 2$, $\rm Ag$-rich material is given by $x = 6$, and whenever the material in the experiment is the hybrid of the two or indistinguishable, we have used the generic chemical formula, ${\rm Ag}_x\rm Ni_2TeO_6$ with defined ranges of $x$ given where possible/relevant. Moreover, whenever there are $\rm Ag$ vacancies present in the average material, we have referred to the material as $\rm Ag$-deficient, and used the chemical formula ${\rm Ag}_{2 - x}\rm Ni_2TeO_6$. With this notation, $\rm Ag_2Ni_2TeO_6$ is monolayered, ${\rm Ag}_x{\rm Ni_2TeO_6}$ with $2 \leq x \leq 7$, as ascertained by scanning transmission electron microscopy 
energy-dispersive X-ray 
spectroscopy (STEM-EDX), is a mixture of monolayered and bilayered domains and $\rm Ag_6Ni_2TeO_6$ is bilayered.}

\section{\label{section: Results} Results}

Given the tendency for silver-containing materials to completely decompose at high temperatures under ambient pressures, conventional solid-state synthetic routes could not be used in this study.\cite{gupta2021} Therefore, silver-based honeycomb layered tellurates encompassing the global compositions ${\rm Ag_2}M_2{\rm TeO_6}$ ($M = \rm Ni, Mg, Co, Cu, Zn$ and $\rm Ni_{0.5}Co_{0.5}$) were synthesised via a low-temperature topochemical ion-exchange reaction, as explicated in the \textbf{\nameref{section: Methods}} section. The elemental concentrations of the ${\rm Ag_2}M_2{\rm TeO_6}$ compositions were confirmed to be in line with the proprietary compositions of ${\rm Ag_2}M_2{\rm TeO_6}$ using inductively coupled plasma atomic emission spectroscopy (ICP-AES), as provided in \textbf{Supplementary Information} (\textbf{Supplementary Table 1}). The stoichiometry and homogeneous elemental distribution of the ${\rm Ag_2}M_2{\rm TeO_6}$ materials were further verified using energy-dispersive X-ray spectroscopy (EDX), as shown in the \textbf{Supplementary Information} (\textbf{Supplementary Figures 2, 3 and 4}).

To ascertain the grain size and morphology of the crystal structures, the as-prepared samples were subjected to scanning electron microscopy (SEM), which revealed a uniform distribution of micrometric-sized particles (\textbf{Supplementary Figures 2, 3 and 4}). The grains were also observed to assume flake-like (lamellar-like) shapes—in character with other layered oxides.\cite{masese2018rechargeable, yoshii2019sulfonylamide} The crystallinity and purity of the samples were ascertained through conventional X-ray diffraction (XRD) analyses as shown in \textbf{Supplementary Figure 5}. From the XRD patterns, no peaks attributed to the initial precursors or impurities were detected, indicating the high purity content of the samples prepared. Even so, the Bragg peaks in the patterns were broad and asymmetric (\textbf{Supplementary Figures 5 and 6a}), making it difficult to precisely validate the crystal structures. Furthermore, some Bragg reflections appeared to merge with the background, ruling out the possibility of accurately modelling the peak shapes.  
In an attempt to obtain a detailed structural characterisation, synchrotron XRD (SXRD) data was obtained from one of the samples; $\rm Ag_2Ni_2TeO_6$ (\textbf{Supplementary Figure 6b}). The material was however found to have undergone SXRD-induced damage, rendering this analytical route inapplicable for the present study.

\begin{figure*}
\centering
\includegraphics[width=0.7\textwidth,clip=true]{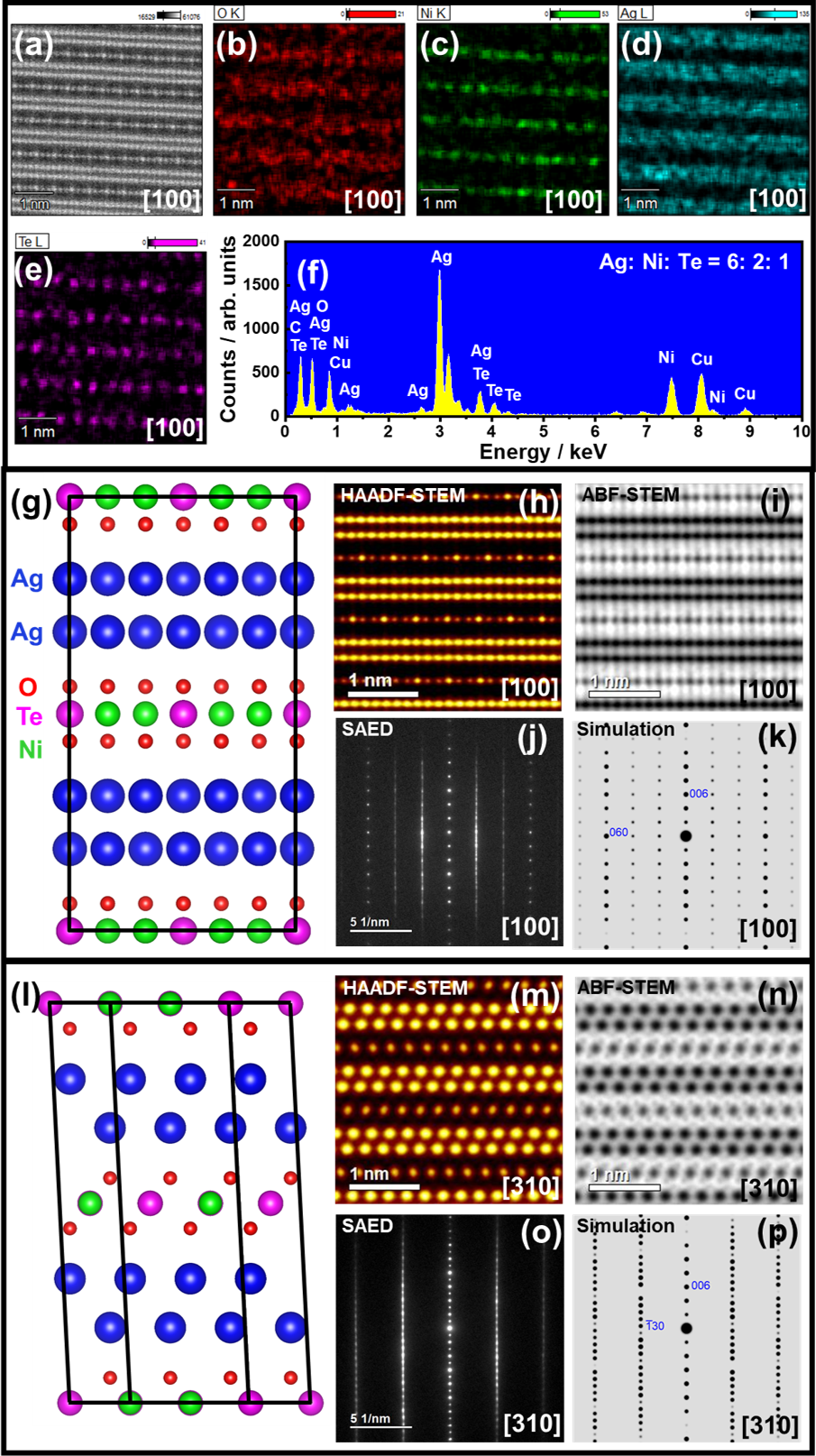}
\end{figure*}
\begin{figure*}
\centering
\caption{\textbf{Figure 1:} Visualisation of bilayered domains of the $\rm Ag$-rich $\rm Ag_6Ni_2TeO_6$ within the global composition, $\rm Ag_2Ni_2TeO_6$. (a) High-angle annular dark-field transmission electron microscopy (HAADF-STEM) images of $\rm Ag_6Ni_2TeO_6$ nanocrystal taken along the [100] zone axis. $\rm Ag$ atom layers (large, dark spots) observed to be sandwiched between slabs of $\rm Te$ and $\rm Ni$ (small, light spots) coordinated with $\rm O$ atoms. (b, c, d, e) Corresponding STEM-energy dispersive spectroscopy (EDS) mapping showing the arrangement of constituent elements ($\rm O, Ni, Ag$ and $\rm Te$). (f) STEM-EDS spectra for a selected section (shown in (a)) confirming the elemental composition of the Ag-rich domain, $\rm Ag_6Ni_2TeO_6$. The $\rm Cu$ and $\rm C$ spectra detected emanated from the TEM holder. (g) Atomistic model of the average structure of $\rm Ag_6Ni_2TeO_6$ derived using STEM analyses along the [100] zone axis. Black lines are used to depict the partial unit cell. (h) HAADF-STEM image, showing $\rm Ag$ atom bilayers sandwiched between slabs comprising $\rm Ni$ and $\rm Te$ atoms. (i) Annular bright-field (ABF) image, affirming the atomic positions of oxygen atoms. (j) Selected area electron diffraction (SAED) patterns taken along the [100] zone axis revealing spot shifts and streaks that suggest the existence of aperiodicity in line with the shifts in the transition metals slabs observed in (h) and (i). (k) Corresponding kinematic simulations based on the structural model shown in (a). (l) Atomistic model of $\rm Ag_6Ni_2TeO_6$ derived using STEM analyses along the [310] zone axis. The black lines are used to depict the partial unit cell. An alternating arrangement in the orientation of oxygen atoms can be seen in the subsequent transition metal atom slabs. (m) HAADF-STEM image, showing $\rm Ag$ atom bilayers sandwiched between slabs containing Te atoms. $\rm Ni$ atoms are superimposed on the position of the $\rm Te$ atoms. (n) Annular bright-field (ABF) image, affirming the alternating orientation of oxygen atoms in successive slabs. (o) Selected area electron diffraction (SAED) patterns taken along the [310] zone axis revealing spot shifts and streaks that suggest the existence of aperiodicity. (p) Corresponding kinematic simulations based on the structural model shown in (l).}\label{Fig_1}
\end{figure*}

It is worth noting that the appearance of significantly broadened peaks in the present ${\rm Ag_2}M_2{\rm TeO_6}$ compositions suggests the existence of defects or disorders in the slab stackings of the layered materials, as has been noted in related honeycomb layered oxides such as $\rm Ag_3Ni_2BiO_6$, $\rm Ag_3Co_2SbO_6$ and $\rm NaKNi_2TeO_6$.\cite{masese2021mixed, berthelot2012new, politaev2010} Therefore, to explicitly visualise the emergent stacking sequences and honeycomb ordering of the ${\rm Ag_2}M_2{\rm TeO_6}$ samples without compromising their structural integrity, aberration-corrected scanning transmission electron microscopy (STEM) was employed, as illustrated in \textbf{Figure 1}. A high-angle annular dark-field (HAADF) STEM image obtained along the [100] zone axis (\textbf{Figure 1a}) shows an array of darker spots ($\rm Ag$ atoms) sandwiched between thinner planes of $\rm Te$- and $\rm Ni$- atom planes (light spots). This atomic arrangement is validated through elemental mapping by STEM-EDX, as shown in \textbf{Figures 1b, 1c, 1d and 1e}. The elemental composition of constituent elements in the observed crystallite domain was further ascertained by STEM-EDX spectra (\textbf{Figure 1f}) to be in the ratio of 6:2:1 for Ag, Ni, and Te, respectively, crucially revealing a Ag-rich $\rm Ag_6Ni_2TeO_6$ crystalline domain. For ease of reference, the contrast ($I$) of the HAADF-STEM image in \textbf{Figure 1a} is proportional to the atomic number ($Z$) of elements and their atomic arrangement (where $I \propto Z^{1.7} \approx Z^2$)\cite {pennycook2006, pennycook1988, pennycook2006materials} as shown in \textbf{Figure 1h}. The image clearly displays a bilayer plane of $\rm Ag$ atoms ($Z = 47$), marked by the larger and brighter golden spots, positioned between the layers of $\rm Te$ atoms ($Z = 52$) denoted by the smaller golden spots, and $\rm Ni$ atoms ($Z = 28$) represented by the darker amber spots. The corresponding annular bright-field (ABF) STEM images (\textbf{Figure 1i}) is obtained to highlight the position of oxygen atoms in the crystal structure. As for ABF-STEM images, $I \propto Z^{1/3}$,\cite {pennycook2006, pennycook1988, pennycook2006materials} which means that elements with lighter atomic mass such as $\rm O$ ($Z = 8$) can be visualised. For a clear visualisation of the structural configuration of the Ag-rich crysalline domain of $\rm Ag_6Ni_2TeO_6$, \textbf{Figure 1g} illustrates a crystal structure model rendered from the STEM images along the [100] zone axis. Here, the $\rm Te-Ni-Ni-Te$ sequential arrangement of the $\rm Te$ atoms and $\rm Ni$ atoms, typical amongst honeycomb structures, is clearly visualised. The atomic arrangements discerned by the STEM analyses were additionally corroborated by selected area electron diffraction (SAED) measurements taken along the [100] zone axis. As shown in \textbf{Figure 1j}, the atoms appear to align in a `streak-like' array of spots {\it in lieu} of distinctly separated spots, indicating the existence of a stacking disorder(s) (fault) across the slab (along the $c$-axis). These results are further validated using kinematically simulated electron diffraction patterns (\textbf{Figure 1k}), which show consistency with the experimentally obtained SAED patterns. The atomic arrangement of atoms when viewed along the [100] zone axis is shown in \textbf{Figure 1l}.

To shed light on the oxygen-atom positions in the crystallite, HAADF- and ABF-STEM images of the $\rm Ag_6Ni_2TeO_6$ nanocrystal were obtained along the [310] zone axis as shown in \textbf{Figures 1m and 1n}, respectively. The oxygen atoms appear to be arranged diagonally in a zig-zag orientation along the $c$-axis—an orientation similar to those registered by the precursor materials prior to the topotactic ion exchange. The atomic arrangement along the [310] zone axis is confirmed using experimentally obtained SAED patterns (\textbf{Figure 1o}) and their corresponding kinematically simulated electron diffraction patterns (\textbf{Figure 1p}). The patterns obtained along the [310] zone axis manifests atomic streaks resembling those derived along the [100] zone axis, further substantiating the existence of stacking variants (faults or disorders) across the slabs. As such, an extensive examination of the stacking sequences is still necessary to garner deeper insight into their crystallographic information.

\begin{figure*}
\centering
\includegraphics[width=0.75\textwidth,clip=true]{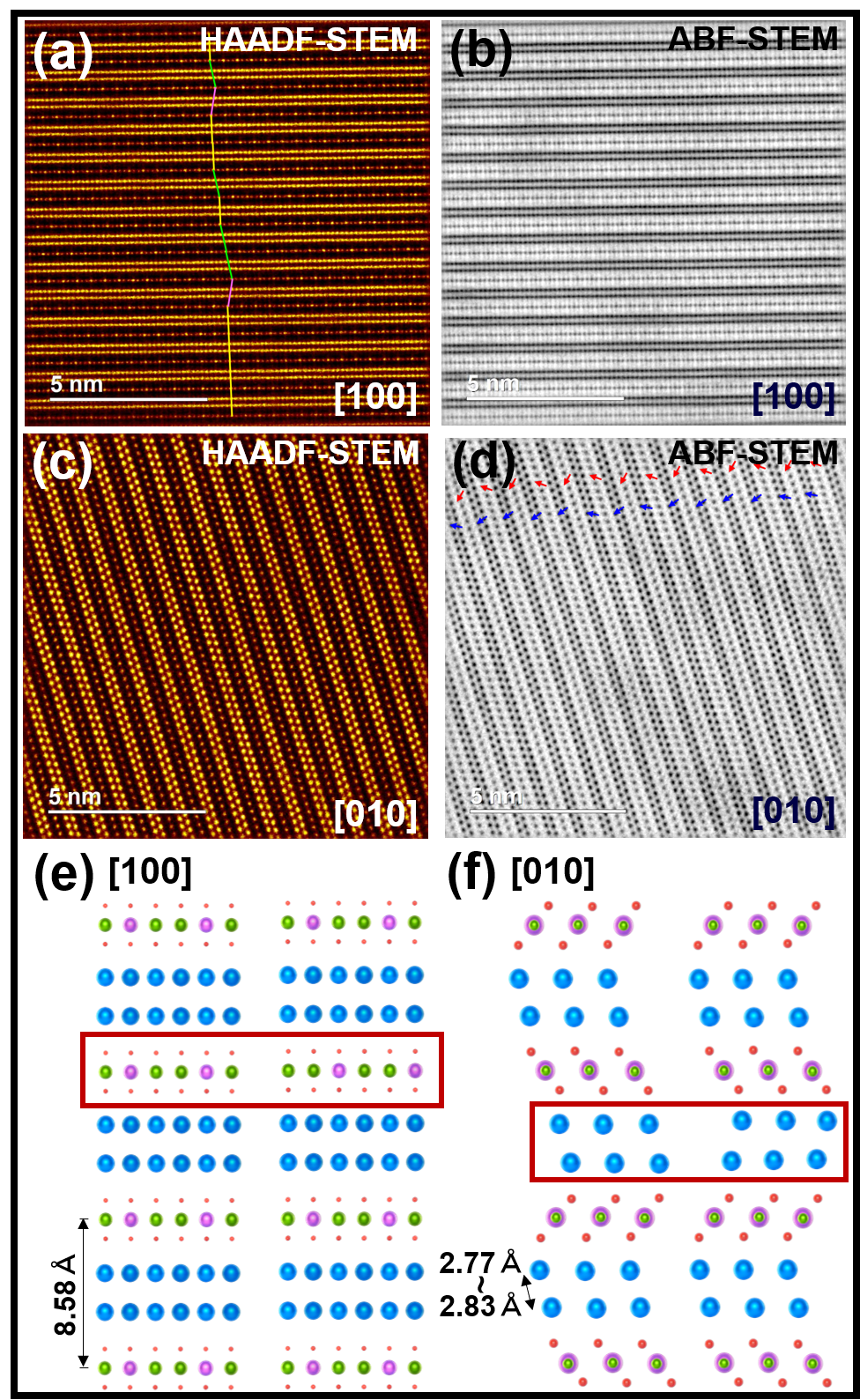}
\end{figure*}
\begin{figure*}
\centering
\caption{\textbf{Figure 2:} High-resolution STEM imaging of $\rm Ag$ atom bilayers along multiple zone axes. (a) HAADF-STEM image of $\rm Ag_6Ni_2TeO_6$ crystallite taken along [100] zone axis showing the aperiodic ordering sequence of Ni and Te atoms in successive slabs and (b) Corresponding ABF-STEM image. (c) Visualisation (along the [010] zone axis) using HAADF-STEM, and (d) Corresponding ABF-STEM image, showing aperiodicity also in the arrangement of $\rm Ag$ in their respective sites across the slabs (or along the $c$-axis). The red arrows highlight a periodic sequence of orientation of oxygen atoms across the slab and the blue arrows show shifts in the alignment of the $\rm Ag$ atom bilayer along the ab plane. (e) Various arrangements of $\rm Ag$ atoms in $\rm Ag_6Ni_2TeO_6$ observed along the $b$-axis. (f) Shifts in the arrangement of $\rm Ni$ and $\rm Te$ atom slabs along the $c$-axis. Silver atoms are shown in blue. Oxygen atoms are shown in red, whereas $\rm Ni$ and $\rm Te$ atoms are shown in green and pink, respectively.}
\label{Fig_2}
\end{figure*}

Accordingly, to ascertain honeycomb ordering and the nature of stacking variations in the crystallite, the samples were subjected to high-magnification STEM analyses, as illustrated in \textbf{Figure 2}. In ordered honeycomb layered tellurate structures, the Te atoms (smaller golden spots) are typically positioned directly below or above the adjacent slabs in idyllic vertical arrays. However, the HAADF-STEM images taken along the [100] zone axis (\textbf{Figure 2a}) reveal that in certain domains, the slabs deviate laterally from the `optimal' arrays (as highlighted by the green and pink lines), indicating the occurrence of stacking faults across the slab stacking direction ($c$-axis). For clarity, the right and left shifts of the $\rm Ni/Te$ atom slabs are denoted by green and pink lines, respectively. The corresponding ABF-STEM image (\textbf{Figure 2b}) further underpin the shifts of Te/Ni atom slabs, albeit not as discernible as that of the HAADF-STEM image. Similar aperiodic shifts in the $\rm Ni/Te$ slabs are also observed in the STEM images taken along the [110] zone axes, as shown in \textbf{Supplementary Figure 7}. The occurrence of multiple disorders involving shifts in the metal slab layers along the $c$-axis not only reflects the diversity of the disorders intrinsic in $\rm Ag_6Ni_2TeO_6$ but may also be envisioned to induce other disorders in the arrangement of $\rm Ag$ atom bilayers. 

To investigate the occurrence of disorders in the $\rm Ag$ atom bilayers, HAADF- and ABF-STEM images were taken along the [010] zone axis (\textbf{Figures 2c and 2d}). Although the orientation of the oxygen atoms across the $\rm Te/Ni$ slabs appears to follow a periodic sequence across the slab (as highlighted by red arrows in \textbf{Figure 2d}), the alignment of the $\rm Ag$ atom bilayer is seen to shift along the ab plane (perpendicular to the $c$-axis) as indicated by the blue arrows. Here, the orientation of adjacent $\rm Ag$ bilayer planes is observed to frequently invert with no periodicity across the slabs, indicating a lack of coherency in their orientation along the $c$-axis. Crystal schematic illustrations are provided in \textbf{Figures 2e and 2f} for an extensive review of the disorders in the arrangement of the transition metal slabs and the orientation of the $\rm Ag$ atom bilayers along the slab. The $\rm Ag_6Ni_2TeO_6$ crystallite appears vastly disordered with no coherence between the stacked transition metal layers and the silver atoms. Similar observations were made across the global honeycomb layered oxide compositions, ${\rm Ag_2}M_2{\rm TeO_6}$ presented in this study (see \textbf{Supplementary Figures 8, 9, 10, 11, 12, 13 and 14}). Indeed, the variegation of structural defects/disorders visualised in the atomic resolution images are far beyond the reach of diffraction measurements, ratifying the need for high-resolution STEM in the exploration of similar layered materials.

\section{\label{section: Discussion} Discussion}

\subsection{Experimental considerations}

Herein, we report for the first time, silver-based honeycomb layered tellurates embodying global compositions, ${\rm Ag_2}M_2{\rm TeO_6}$ (where $M = 3d$ transition metals or $s$-block elements such as $\rm Mg$) synthesised via topochemical ion-exchange. Atomic-resolution STEM analyses conducted along multiple zone axes reveal these tellurates, \textit{i.e.}, $\rm Ag_2Ni_2TeO_6$, $\rm Ag_2Mg_2TeO_6$, $\rm Ag_2Co_2TeO_6$, $\rm Ag_2Cu_2TeO_6$ and $\rm Ag_2NiCoTeO_6$ (\textbf{Figure 1} and see \textbf{Supplementary Figures 8, 9, 10, 11, 12, 13 and 14}), to predominantly encompass, within their $\rm Ag$-rich domains (\textit{i.e.}, $\rm Ag_6Ni_2TeO_6$, $\rm Ag_6Mg_2TeO_6$, $\rm Ag_6Co_2TeO_6$, $\rm Ag_6Cu_2TeO_6$ and $\rm Ag_6NiCoTeO_6$), silver atom bilayers interspersed between honeycomb slabs. These $\rm Ag$-atom bilayered tellurates were observed to engender crystallites with significantly larger interlayer distances and variegated structural disorders—attributes poised to propagate fascinating two-dimensional interactions, phase transitions and rapid cation diffusion within the materials. Amongst the numerous crystallites investigated, the global composition $\rm Ag_2Ni_2TeO_6$ exhibits $\rm Ag$-rich crystalline domains with interslab distances of $\sim$ 9 \AA\, with a manifold of structural disorders (\textbf{Figure 2}), making it the exemplar material of focus for this study. It is worth mentioning that to date, bilayered structures entailing $\rm Ag$-atoms have not been reported amongst honeycomb layered oxides, despite the rich structural diversity manifested by these materials. In fact, STEM analyses conducted on a bismuthate analogue, $\rm Ag_3Ni_2BiO_6$, prepared using the present synthesis protocols, demonstrate the crystallites to have a monolayered arrangement of monovalent $\rm Ag$ atoms with significantly smaller interlayer distances (see \textbf{Supplementary Figures 15, 16 and 17}). Moreover, a previous simulation result of Ag-based honeycomb layered oxides\cite{tada2022implications} using the Kohn-Sham formalism\cite{kohn1965self} found only monolayered frameworks, \red{indicative of the 
challenges and controversy faced by conventional and non-conventional bonding schemes to effectively reproduce Ag-based 
structures.\cite{lobato2021comment, yin2021reply, vegas2020re}} As such, this study not only represents a major milestone in the exploration of honeycomb layered oxides functionalities but also expounds on the structural expedience of honeycomb layered tellurates (\textbf{Supplementary Figure 18}).

The material knowledgebase for compounds manifesting $\rm Ag$-atom bilayered structures remains limited.\cite{johannes2007formation, yoshida2020static, taniguchi2020butterfly, yoshida2011novel, matsuda2012partially, yoshida2008unique, yoshida2006spin} Thus, their occurrence in the present honeycomb layered tellurates betoken significant progress in the advancement of the crystal structural versatility of honeycomb layered oxides. Until now, the advancement of compounds manifesting Ag-atom bilayers has been heavily curtailed by their stringent synthetic conditions, which typically involve giga-Pascal scale pressures and synthesising precursors under elevated oxygen pressures and temperatures.\cite{schreyer2002synthesis, yoshida2020static, taniguchi2020butterfly} As an alternative route, this study employs a high molar silver salt-to-precursor ratio to develop these bilayered structures via a low-temperature metathetic (topochemical ion-exchange) synthetic route. Equivalent molar ratios of initial precursors and $\rm AgNO_3$ molten salt in the case of $\rm Ag_2Ni_2TeO_6$ — were found to be insufficient in facilitating a complete $\rm Ag^+$ ion exchange (see \textbf{Supplementary Figure 19}). Although the resulting crystallites formed are predominantly $\rm Ag$-atom bilayered structures, defects in the arrangement of silver atoms were exhibited in some crystallites, characterised by the presence of $\rm Ag$-deficient domains with atoms in their amorphous state (single $\rm Ag$ atom layers) alongside $\rm Ag$-rich domains with $\rm Ag$-atom bilayers (\textbf{Supplementary Figures 20, 21, 22 and 23}). This postulation was verified by TEM-EDX measurements, which demonstrate ${\rm Ag_2}M_2{\rm TeO_6}$ compositions to have a rich global composition comprising $\rm Ag$-rich and $\rm Ag$-deficient regimes in close proximity. 

\begin{figure*}
\centering
\includegraphics[width=0.75\textwidth,clip=true]{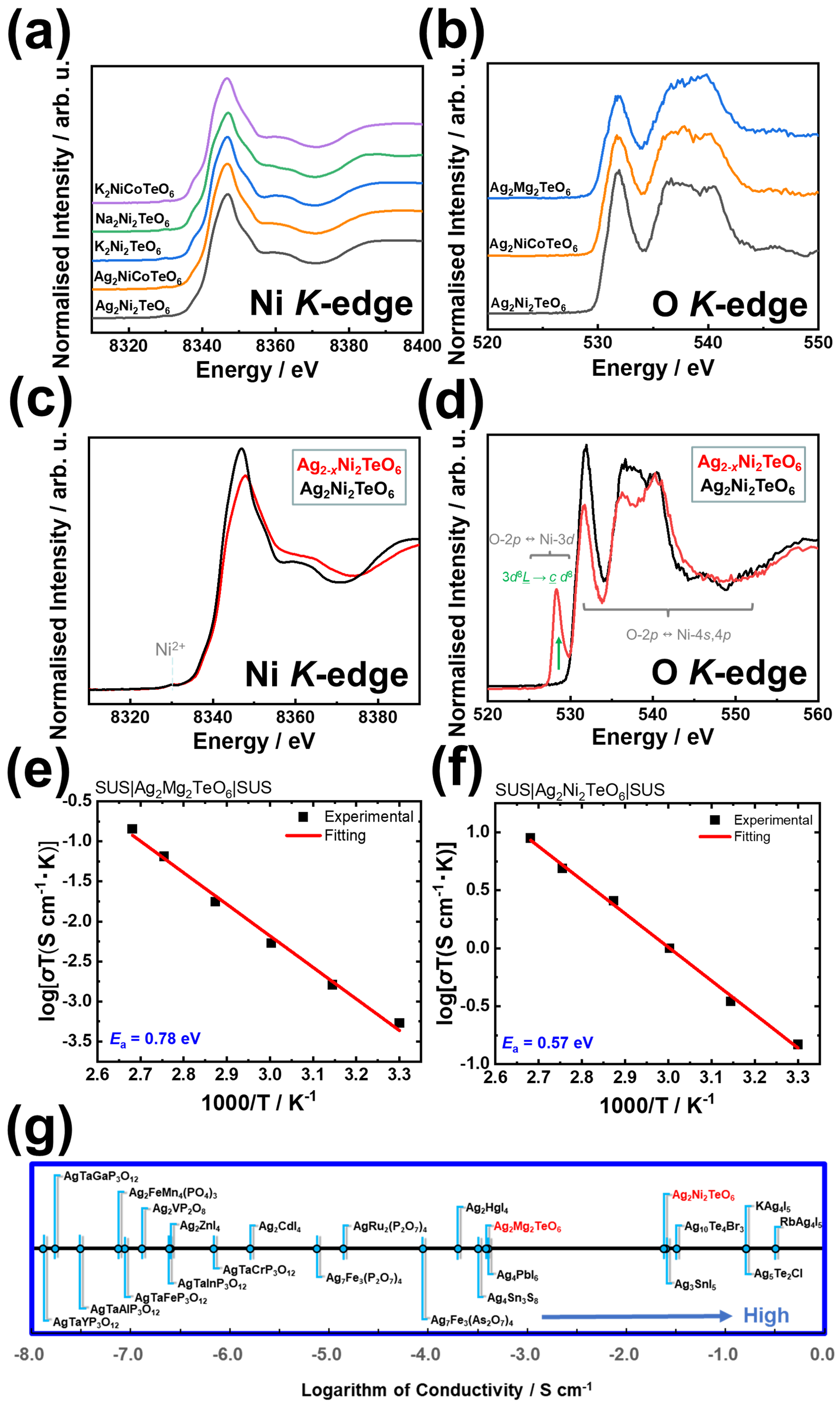}
\end{figure*}
\begin{figure*}
\centering
\caption{\textbf{Figure 3:} Spectroscopic measurements of the global compositions, ${\rm Ag_2}M_2{\rm TeO_6}$ ($M = \rm Mg$ and $\rm Ni$). (a) Normalised $\rm Ni$ $K$-edge {\it ex situ} X-ray absorption spectra (XAS) of $\rm Ag_2Ni_2TeO_6$ and related derivatives ($\rm Ag_2NiCoTeO_6$ and $\rm Ag_2Mg_2TeO_6$) collected along with reference $\rm Ni^{2+}$ compounds ($\rm Na_2Ni_2TeO_6$, $\rm K_2Ni_2TeO_6$ and $\rm K_2NiCoTeO_6$) and (b) Normalised O $K$-edge {\it ex situ} XAS spectra of ${\rm Ag_2}M_2{\rm TeO_6}$ ($M = \rm Mg, Ni$ and $\rm Ni_{0.5}Co_{0.5}$) taken in fluorescence yield (FY) mode—highly sensitive to the innate bulk properties. (c) Normalised Ni $K$-edge {\it ex situ} XAS spectra of $\rm Ag_2Ni_2TeO_6$ pristine electrode and charged electrode (${\rm Ag}_{2-x}\rm Ni_2TeO_6$). (d) Normalised O $K$-edge {\it ex situ} XAS spectra of $\rm Ag_2Ni_2TeO_6$ pristine electrode and charged electrode (${\rm Ag}_{2-x}\rm Ni_2TeO_6$). Ligand holes are created in the ${\rm O}\,\,2p$ bands during $\rm Ag$-ion extraction. (e) Arrhenius plots derived from electrochemical impedance spectroscopy (EIS) measurements of $\rm Ag_2Mg_2TeO_6$ and (f) $\rm Ag_2Ni_2TeO_6$. (g) Comparative plots of the ionic conductivity values attained in representative $\rm Ag$-based ionic conductors reported along with the honeycomb layered tellurates.\cite{hull2000structural, hull2001structural, hull2002crystal, hull2002structural, nilges2004structure, nilges2005structures, matsunaga2004structural, hull2005ag+, lange2006ag10te4br3, lange2007polymorphism, angenault1989conductivite, rao2005preparation, daidouh2002structural, daidouh1997structure, fukuoka2003crystal, quarez2009crystal}}\label{Fig_3}
\end{figure*}

From a chemical perspective, the formation of Ag-atom bilayered structures (as shown in \textbf{Figures 2 and 3}) can be attributed to the tendency of $\rm Ag$ to assume anomalous valency states (also referred to as sub-valent states) such as oxidation states between $0$ and $1+$. As a result, the $\rm Ag$ ions aggregate to form atomic coordinations ({\it i.e.}, $\rm Ag-Ag'$ bonds) resemblant of silver metal (metallic silver) topologies\cite{johannes2007formation, beesk1981x, ahlert2003ag13oso6, kovalevskiy2020uncommon, kohler1985electrical}, which in principle violate conventional bonding mechanisms and electronic structures. In general, the constituent elements of the present global compositions can be assigned to the valency states of $\rm Ni^{2+}$, $\rm Ag^{1+}$ and $\rm Te^{6+}$ to yield a valency description of $\rm Ag_2^{1+}Ni_2^{2+}Te^{6+}O_6^{2-}$. To ascertain these valency states, X-ray photoelectron spectra (XPS) of the $\rm Ag_2Ni_2TeO_6$ crystallite and its related derivatives ($\rm Ag_2NiCoTeO_6$ and $\rm Ag_2Mg_2TeO_6$) were obtained at the binding energies of ${\rm Ag}\,\, 3d$, ${\rm Ni}\,\,2p$ and ${\rm Te}\,\,3d$, as provided in the \textbf{Supplementary Figures 24 and 25}. Since the measurements were performed on the global compositions, the existence of $\rm Ag$ sub-valent states within the $\rm Ag$-rich domains could not be unequivocally established using spectroscopic techniques such as XPS and XAS. In particular, since the as-prepared material with a global composition of $\rm Ag_2Ni_2TeO_6$ (as ascertained by ICP-AES) has an inseparable mixture of Ag-deficient phases (such as ${\rm Ag}_{2 - x}\rm Ni_2TeO_6$, $0 < x < 2$) and Ag-rich (expected) sub-valent phases such as $\rm Ag_6Ni_2TeO_6$, this necessitates one to perform STEM electron energy-loss spectroscopy (EELS) on Ag-rich nanocrystallite domains at the Ag {\it M}-, Te {\it M}-, Ni {\it L}- and O {\it K}-edges in order to distinguish these phases, which proved challenging to conclusively perform in the present work. Meanwhile, the divalent nature of $\rm Ni$ ($\rm Ni^{2+}$) is further corroborated through X-ray absorption spectroscopy (XAS) performed on the global composition, $\rm Ag_2Ni_2TeO_6$ and its related derivatives ($\rm Ag_2NiCoTeO_6$ and $\rm Ag_2Mg_2TeO_6$) at the $\rm Ni$ $K$-edge, as shown in \textbf{Figure 3a}. Further, O $K$-edge XAS spectra of $\rm Ag_2Ni_2TeO_6$ along with related tellurate compositions are taken in the bulk-sensitive fluorescence yield mode to establish the valency of $\rm O$ atoms (\textbf{Figure 3b}). Here, no spectral features ascribed to oxygen hole formation were identified, indicating that the valency of oxygen does not contribute to the formation of the structures observed. Thus, since the valency states of the other metal elements in the $\rm Ag$-rich domains were ascertained to be $\rm Ni^{2+}$ (divalent) for $\rm Ni$ atoms, and mixed valency states of $\rm Te^{4+}$ and $\rm Te^{6+}$ for $\rm Te$ atoms, and the general chemical formula can be confirmed by the XPS and XAS results alongside the STEM-EDX spectra to be given by ${\rm Ag_6Ni_2TeO_6} = {\rm Ag}_6^{(1 - \delta)+}{\rm Ni_2^{2+}Te}_x^{4+}{\rm Te}_y^{6+}\rm O_6^{2-}$ ($x \geq 0$, $y \geq 0$, $x + y = 1$), $\rm Ag$ must be sub-valent by the charge neutrality requirement (\textit{i.e.} $6\times (1 - \delta) + 2\times 2 + 4x + 6y + 6\times (2-) = 0 \rightarrow 2x + 3y = 1 + 3\delta$, together with $x + y = 1$, are simultaneous equations which can be solved to yield, $\delta = (y + 1)/3$ thus giving the $\rm Ag$ sub-valency range, $1/3+ \leq (1 - \delta)+ \leq 2/3+$ for $0 \leq y \leq 1$). 

Whilst we have successfully shown, by the charge neutrality argument constrained by the experimental data, that the $\rm Ag$-rich domain of the present material must comprise sub-valent $\rm Ag$ cations, we cannot further determine the exact value of this sub-valency which ought to lie within the bound, $1/3+ \leq (1 - \delta)+ \leq 2/3+$. Meanwhile, first-principles computations (\textbf{Supplementary Figure 26 and Supplementary Note 1})  
suggest a complex valency composition encompassing an admixture of multiple valency states such as $\rm Ag^{1/3+}_6Ni^{2+}_2Te^{6+}O_6^{2-}$, $\rm Ag^{2/3+}_6Ni^{2+}_2Te^{4+}O_6^{2-}$, $\rm Ag^{1/2+}_6Ni^{2+}_2Te^{4+}_{1/2}Te^{6+}_{1/2}O_6^{2-}$ and $\rm Ag^{1/2+}_4Ni^{2+}_2Te^{6+}O_6^{2-}$, which evince to the contribution of sub-valent states of $\rm Ag$ atoms in the formation of silver atom bilayers.\cite{johannes2007formation, yoshida2020static, taniguchi2020butterfly, yoshida2011novel, matsuda2012partially, yoshida2008unique, yoshida2006spin} The snippets of structural information gathered in this study altogether allude to the rich global composition of the present tellurates, which entail $\rm Ag$-deficient domains with a valency description of ${\rm Ag^{1+}}_{2-x}{\rm Ni}_2^{(2+z/2)+}{\rm Te}^{(6 - y)+}\rm O_6^{2-}$ ($z = x + y$, $0 \leq x < 2$\red{, $0 \leq y \leq 2$}) alongside $\rm Ag$-rich domains comprising an admixture of $\rm Ag^{1/2+}_4Ni^{2+}_2Te^{6+}O_6^{2-}$, $\rm Ag^{1/2+}_6Ni^{2+}_2Te^{4+}_{1/2}Te^{6+}_{1/2}O_6^{2-}$ and $\rm Ag^{2/3+}_6Ni^{2+}_2Te^{4+}O_6^{2-}$, \textit{etc}. with varied atomic occupancies. In fact, such sub-valent $\rm Ag$ bilayered frameworks have hitherto been reported in materials such as $\rm Ag_2^{1/2+}F^{1-}$, $\rm Ag_3^{2/3+}O^{2-}$, $\rm Ag_2^{1/2+}Ni^{3+}O_2^{2-}$ and hybrids (alternating monolayers and bilayers) such as $\rm Ag_3^{2/3+}Ni_2^{3+}O_4^{2-}$.\cite{schreyer2002synthesis, yoshida2006spin, beesk1981x, johannes2007formation, sorgel2007ag3ni2o4} A crucial observation is that, these frameworks not only involve sub-valent states ($1/2+$ or $2/3+$ \textit{etc}.) but also their $\rm Ag$ bilayers are formed by two triangular lattices of an apparent bifurcated bipartite honeycomb lattice. Meanwhile, in the case where there is a monolayered counterpart, \textit{e.g.} $\rm Ag^{1+}Ni^{3+}O_2^{2-}$, the $\rm Ag$ is not only monovalent but also the lattice is triangular and monolayered.\cite{schreyer2002synthesis, yoshida2006spin, johannes2007formation, sorgel2007ag3ni2o4} 
Indeed, this is the case for $\rm Ag_3^{1+}Ni_2^{2+}Bi^{5+}O_6^{2-}$, whereby the $\rm Ag$ atom is monovalent, and the lattice is not only monolayered but triangular\cite{berthelot2012new}, in contrast to the present material reported herein. Meanwhile, the $\rm Ag$ sub-valent state $1/3+$ remains conspicuously absent in reported layered materials\cite{schreyer2002synthesis, yoshida2006spin, johannes2007formation} and other silver-rich oxides\cite{wang2018trapping, haraguchi2021formation, derzsi2021ag, kovalevskiy2020uncommon, ahlert2003ag13oso6, jansen1992ag5geo4, jansen1990ag5pb2o6, argay1966redetermination, beesk1981x, bystrom1950crystal}, despite its presence reported \textit{e.g.} in $\rm Ag_3^{1/3+}O^{2-}H^{1+}$.\cite{molleman2015surface} Finally, although the sub-valency of silver remains unascertained in the present material via direct XPS measurements of $\rm Ag$ binding energies, the fact that the $\rm Ag$-rich domains are bilayered with an apparent bifurcated bipartite honeycomb lattice demonstrates that such domains fit well within the aforementioned class of bilayered sub-valent $\rm Ag$-based frameworks, thus corroborating the charge neutrality argument provided earlier for the sub-valency of $\rm Ag$. Further information on the sub-valent nature of Ag in the bilayered domains could be garnered from a direct visualisation of the local coordination of Ag atoms using high-resolution STEM (\textbf{Figure 1}). The shortest $\rm Ag-Ag'$ distances in the bilayered $\rm Ag_6Ni_2TeO_6$ domains were found to be those between adjacent Ag atoms of subsequent layers (in the ranges of 2.77 \AA $\sim$ 2.83 \AA \,\, (\textit{viz}., 2.80 $\pm$ 0.03 \AA)), as highlighted in \textbf{Figure 3f} and \textbf{Supplementary Figure 27}. These $\rm Ag-Ag'$ distances of $\rm Ag_6Ni_2TeO_6$ are akin to those of sub-valent $\rm Ag_2^{1/2+}NiO_2$ (2.836 \AA) and $\rm Ag_2^{1/2+}F$ (2.814 \AA),\cite {schreyer2002synthesis, argay1966redetermination} suggesting a universal $\rm Ag - Ag'$ bonding mechanism for such bilayered and other related materials. For instance, this universality beyond layered materials is also exhibited by the bifurcated $\rm Ag$ honeycomb structure in $\rm Ag_{16}^{1/2+}B_4O_{10}$ (constituting a tetrahedral shape conjectured by the authors of \cite{kovalevskiy2020uncommon} to habour excess localised electrons responsible for the reported $\rm Ag$ sub-valency of $1/2+$), with a bond length of order $\sim 2.8$ \AA.\cite{kovalevskiy2020uncommon}

Considering the presence of mobile $\rm Ag$ cations sandwiched between transition metal slabs comprising highly electronegative $\rm Ni^{2+}$, empirical insight into the possibility of electrochemical extraction of $\rm Ag$-ions from the $\rm Ag_2Ni_2TeO_6$ structures would be invaluable in their future utility. Thus, the electrochemical performance of 
$\rm Ag_2Ni_2TeO_6$ electrode was investigated through cyclic voltammetry conducted on $\rm Ag$ half-cells, as detailed in the \textbf{\nameref{section: Methods}} section. The voltammograms obtained illustrate the occurrence of an oxidative peak at around 1.4 V versus $\rm Ag^{+}$/$\rm Ag$, pointing to the prospects of silver-ion extraction at high voltages (see \textbf{Supplementary Figure 28}). However, no reduction peaks were observed, suggesting the occurrence of an irreversible phase transformation or structural deterioration (amorphisation)—which was further affirmed by the corresponding galvanostatic cycling tests. The amorphisation/phase transformation of $\rm Ag_2Ni_2TeO_6$ is further evident in {\it ex situ} XRD measurements as shown in \textbf{Supplementary Figure 29}. To investigate the atomistic mechanisms governing the silver-ion extraction process in $\rm Ag_2Ni_2TeO_6$, {\it ex situ} XAS spectra were obtained from pristine and charged $\rm Ag_2Ni_2TeO_6$ electrodes at the $\rm Ni$ $K$- and $\rm O$ $K$-edges. As shown in \textbf{Figure 3c}, no significant changes in the spectral features of the electrode are observed during silver-ion extraction, an indication that $\rm Ni$ predominantly remains in the divalent state throughout the process. On the other hand, the $\rm O$ $K$-edge XAS spectra (\textbf{Figure 3d}) displays a sharp increase in the intensity of the pre-edge peak centred around 528 eV during the charging process. This observation evinces that the extraction of silver ions from $\rm Ag_2Ni_2TeO_6$ is accompanied by a rapid formation of oxygen ligand holes.

In principle, $\rm Ag$ ion extraction can be rationalised to increase the valency state of nickel from $\rm Ni^{2+}$ to $\rm Ni^{3+}$, where $\rm Ni^{3+}$ has a predominant electronic ground state of $3d^7$. However, the agitations in the Ni-atom electronic configurations triggered by their hybridisation with $\rm O$-$2p$ orbitals engender a ground state characterised by the $3d^8\underline{L}$ orbital character (for clarity, $\underline{L}$ denotes the ligand hole)—akin to those observed in the charged states of compounds such as $\rm LiNiO_2$.\cite{uchimoto2001changes} Accordingly, the increased intensity observed in the pre-edge peak at 528 eV during charging ($\rm Ag$ ion extraction) can be attributed to the transition into the $3d^8\underline{L}$ ground state. Thus, the core transitions during this process can be assigned as $3d^8\underline{L}\rightarrow \underline{c}d^8$. It is worth pointing out that although honeycomb layered oxides such as $\rm Li_4FeSbO_6$ have been shown to exhibit reversible oxygen-redox capabilities,\cite {mccalla2015} the present spectroscopic and diffraction measurements indicate the rapid formation of oxygen holes that debilitate the structural integrity of $\rm Ag_2Ni_2TeO_6$ during silver-ion extraction at high voltages.

\textit{Nota bene}, the prominence gained by honeycomb layered oxides has to some extent been banked on the high voltage capabilities and fast ionic conductivities seen in materials such as $\rm Na_2Ni_2TeO_6$ and $\rm Na_2Mg_2TeO_6$.\cite{kanyolo2021honeycomb, kanyolo2022advances, evstigneeva2011new, li2018new, wu2018sodium} Therefore, investigating the ionic conductivities of their silver analogues (\textit{i.e.}, $\rm Ag_2Ni_2TeO_6$ and $\rm Ag_2Mg_2TeO_6$) under various temperature conditions would be an integral step in determining their innate capabilities. The compounds were subjected to thermal gravimetric analyses to ascertain their thermal stability (\textbf{Supplementary Figures 30 and 31}). Subsequently, their ionic conductivities at different temperatures were assessed, as shown by the Arrhenius plots in \textbf{Figures 3e and 3f}. Detailed experimental protocols are provided in the \textbf{\nameref{section: Methods}} section. In the temperature range of 30-100 $^{\circ}$C, $\rm Ag_2Ni_2TeO_6$ (with a pellet compactness of $\sim$ 84 \%) was determined to have an activation energy of about 0.57 eV, which was calculated by fitting the alternating current data with the Arrhenius equation. The conductivity of the material, which predominantly emanates from ionic diffusion, was found to be $4.88 \times 10^{-4} \,\,\rm S\,cm^{-1}$ at 30 $^{\circ}$C and $2.39 \times 10^{-2} \,\,\rm S\,cm^{-1}$ at 100 $^{\circ}$C. On the other hand, $\rm Ag_2Mg_2TeO_6$ (with a pellet compactness of $\sim$74 \%) displays a predominant ionic conductivity of $1.77 \times 10^{-6} \,\,\rm S\,cm^{-1}$ at 30 $^{\circ}$C and $3.84 \times 10^{-4} \,\,\rm S\,cm^{-1}$ at 100 $^{\circ}$C. For comparison, the bulk ionic conductivities of the silver-based tellurates are presented alongside other reported silver-ion superionic conductors in \textbf{Figure 3g}.\cite{hull2000structural, hull2001structural, hull2002crystal, hull2002structural, nilges2004structure, nilges2005structures, matsunaga2004structural, hull2005ag+, lange2006ag10te4br3, lange2007polymorphism, angenault1989conductivite, rao2005preparation, daidouh2002structural, daidouh1997structure, fukuoka2003crystal, quarez2009crystal} Until now, binary and ternary silver chalcogenides, silver chalcogenidehalides and silver polychalcogenides have dominated the list of materials with fast $\rm Ag$-ion conduction.\cite{hull2000structural, hull2001structural, hull2002crystal, hull2002structural, nilges2004structure, nilges2005structures, matsunaga2004structural, hull2005ag+, lange2006ag10te4br3, lange2007polymorphism} However, from these ion conductivity plots, it is apparent that the present class of honeycomb layered tellurates (with a global composition of ${\rm Ag_2}M_2{\rm TeO_6}$) confers relatively higher $\rm Ag$-ion conductivity in comparison. It is essential to highlight that $\rm Ag$ ion-based layered oxide materials with high ionic conductivities have yet to be reported. Therefore, these results unveil new prospects of utilising ${\rm Ag_2}M_2{\rm TeO_6}$ honeycomb layered oxides compositions as feasible solid electrolytes for electrochemical devices such as all-solid-state $\rm Ag$-ion batteries.\cite {takada1990, guo2006, inoishi2018, kirshenbaum2016, glukhov2022, delaizir2012}

\begin{figure*}
\centering
\includegraphics[width=0.95\textwidth,clip=true]{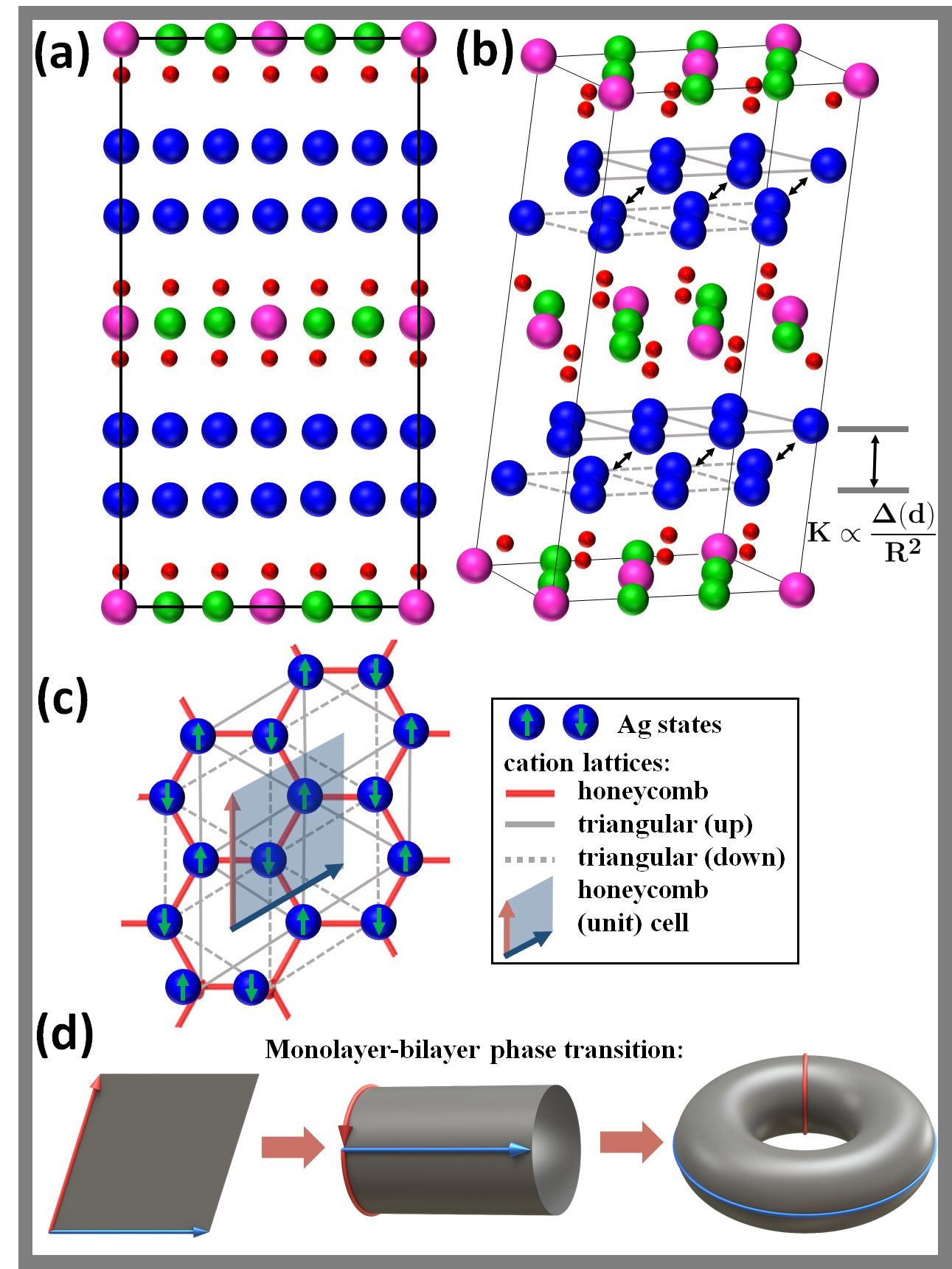}
\end{figure*}
\begin{figure*}
\centering
\caption{\textbf{Figure 4:} Structure, geometry and topological features in the silver-rich honeycomb layered tellurate: ${\rm Ag_6}M_2{\rm TeO_6}$ ($\rm Ag$ atoms are drawn in blue, $\rm Te$ atoms in pink, $\rm Ni$ atoms in green and O atoms in red) (a) A unit cell of ${\rm Ag_6}M_2{\rm TeO_6}$ showing the alignment of the atoms as viewed in the [100] direction. (b) A perspective view of the unit cell of ${\rm Ag_6}M_2{\rm TeO_6}$ shows the alignment of the atoms. The $\rm Ag$ layers form two triangular lattices (drawn as continuous grey lines (up) or dashed grey lines (down)) comprising a single bilayer. The bilayers are separated a distance, $R$ by a stabilising energy gap, $K \propto \Delta(d)/R^2$ arising from SU($2$)$\times$U($1$) spontaneous symmetry breaking (finite $\rm Ag-Ag'$ argentophilic interactions, \textbf{Supplementary Information, Supplementary Note 2}). (c) The honeycomb bilayer lattice of $\rm Ag$ atoms in ${\rm Ag_6}M_2{\rm TeO_6}$ drawn as red lines as viewed from the [001] direction, showing the honeycomb unit cell (the transparent grey rhombus with unit vectors drawn as black and red arrows) and the triangular lattices described in (b). Each $\rm Ag$ atom in the honeycomb unit cell is assigned a pseudo-spin up or down (drawn as green arrows) with opposite orientation to reflect the $S$ transformation on the honeycomb lattice, hence introducing two chiral states of $\rm Ag$ in the honeycomb lattice, $\Lambda$. (d) The topology of the honeycomb unit cell (flat torus, $\Lambda \in \mathbb{R}^2$) depicted in (c) as a torus. The opposite ends of the unit cell are identified with each other, reflecting the translation symmetries ($T$ transformations) of the honeycomb lattice, hence relating the honeycomb lattice to the two-torus (genus one 2D surface, $\mathbb{T}^2$). Our model requires the finite curvature introduced in the direction of the red arrows to be equivalent to a phase transition from 2D to 3D ($\Delta(d) = (d - 2)/2$, $d = 2, 3$) due to $\rm Ag-Ag'$ argentophilic interactions, which is responsible for the energy gap (and hence the bilayers) encountered in (a) and (b).}\label{Fig_4}
\end{figure*}

\subsection{Theoretical considerations}

From a pedagogical perspective, the honeycomb layered ${\rm Ag_2}M_2{\rm TeO_6}$ with $\rm Ag$-rich bilayered domains (\textbf{Figure 4a}) present a prolific playground to investigate the physical origins of argentophilic interactions in such frameworks. Although the $\rm Ag$ sub-valent state is considered integral in the formation of stable bilayers, no apparent mechanism consociating sub-valency to the presence of bilayers has been availed in literature hitherto. Thus, it is prudent to investigate the selection mechanism for the bilayer arrangement supplanting the single layers in other conventional layered materials. Notably, the triangular lattice observed in the $\rm Ag$ bilayers (such as in \textbf{Figure 4b} and \textbf{Figure 4c}) can theoretically be understood as the general manifestation of the underlying emergent geometric field theories associated with the crystalline parameters favoured by the $\rm Ag$ atoms as shown in \textbf{Figure 4(d)}. 

In particular, we consider the crucial features in such bilayered frameworks to be: (i) the unconventional $\rm Ag-Ag'$ bonding between like charges, whose nature can be interpreted as the already reported argentophilic interaction;\cite {schmidbaur2015, jansen1987} (ii) The existence of $\rm Ag$ sub-valent states in almost all reported bilayered frameworks;\cite {schreyer2002synthesis, matsuda2012partially, ji2010, yoshida2020static, yoshida2011novel, yoshida2008unique, yoshida2006spin, argay1966redetermination, beesk1981x} (iii) the apparent bifurcated bipartite honeycomb lattice. Fairly recently, an idealised model of cations describing the diffusion of cations in monolayered frameworks was formulated, whereby the number of vacancies created by diffusing cations can be related to the Gaussian curvature by the Poincar\'{e}-Hopf theorem for an emergent geometric theory consistent with Liouville conformal field theory (two-dimensional (2D) quantum gravity).\cite{kanyolo2022cationic, kanyolo2020idealised, kanyolo2021honeycomb, kanyolo2021partition, kanyolo2022conformal} It is thus imperative to also reproduce the conclusions of the idealised model along the way, which not only classifies the symmetries of the hexagonal and honeycomb cationic lattices in applicable honeycomb layered oxides but also the topological diffusion aspects in 2D. Indeed, this has been achieved by the theoretical model below.

\subsection{Theoretical model for bilayered honeycomb frameworks}

We shall set Planck's constant, the speed of electromagnetic waves in the material and Boltzmann's constant to unity (respectively, $\hbar = \overline{c} = k_{\rm B} = 1$) and employ Einstein summation convention for all raised and lowered indices unless explicitly stated otherwise.

Due to the electrostatic shielding of electric charge of the nucleus and other factors, electron occupation of orbital energy levels for transition metals can disobey Aufbau principle typically employed in standard chemistry to determine electronic configurations of atoms and their valencies.\cite{schwarz2010full} For group 11 elements, the $nd^{10}$ and $(n + 1)s$ orbitals are at close proximity ($< 3.5$ eV) to each other\cite{blades2017evolution}, $sd$ hybridisation is plausible, and can be responsible for degenerate states. In the case of the neutral $\rm Ag$ atom, the electronic configuration can either be $4d^{10}5s^1$ (labelled as $\rm Ag_{+1/2}$) which yields oxidation state, $4d^{10}5s^0$ given by $\rm Ag^{1+}$, or $4d^95s^2$ (labelled as $\rm Ag_{-1/2}$) which yields oxidation state $4d^{10}5s^2$ given by $\rm Ag^{1-}$, whereby the superscript on $\rm Ag$ denotes the oxidation/valency state (\textit{i.e.,} number of electrons that can readily be lost to achieve stability in a chemical reaction) and the subscript $\pm 1/2$ is a spin degree of freedom transforming under SU($2$) gauge group known as isospin.\cite{yang1954conservation} It is clear that, due to $sd$ hybridisation, $4d^95s^2$ and $4d^{10}5s^1$ will be degenerate. Moreover, due to the doubly occupied $s$ orbital of $4d^95s^2$, another oxidation state, $4d^95s^0$ exits, given by $\rm Ag_0 \rightarrow Ag^{2+}$, implying that the neutral $\rm Ag$ atom is three-fold degenerate. Note that, $\rm Ag_{-1/2}$ and $\rm Ag_0$ are degenerate with essentially the same electronic state, $4d^95s^2$, \textit{albeit} different predisposition to form oxidation states $\rm Ag^{1-}$ and $\rm Ag^{2+}$ respectively. Evidently, such degeneracies must be lifted by introducing symmetry breaking in order to create the appropriate oxidation states stabilising argentophillic bonds in layered materials. In particular, honeycomb layered materials with a monolayered structure tend to either have prismatic or linear coordinations of $\rm Ag$ to oxygen atoms\cite{tada2022implications}, which should result in crystal field splitting of the $4d$ orbitals, whereby $4d_{z^2}$ is the lowest energy level in prismatic coordinations (or the highest energy level in linear coordinations).\cite{burns1993mineralogical, ballhausen1963introduction}

\begin{figure*}
\centering
\includegraphics[width=0.65\textwidth,clip=true]{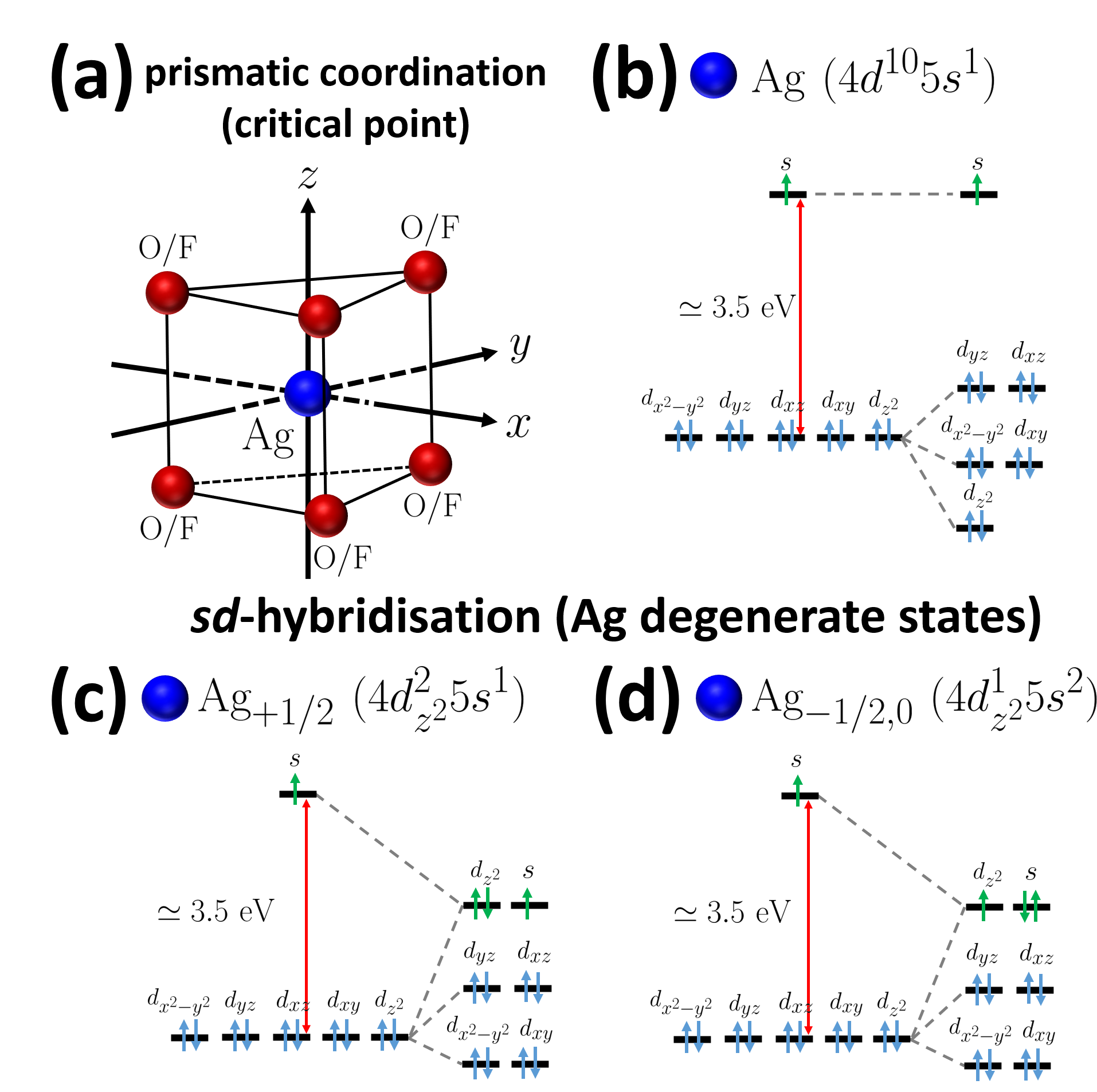}
\end{figure*}
\begin{figure*}
\centering
\caption*{\red{\textbf{Figure 5:} Example of crystal field splitting of $\rm Ag$ atom in a prismatic environment and the ensuing $sd$-hybridisation. (a) An example of expected $sd$-hybridisation in silver, $\rm Ag$ atoms (blue) prismatically coordinated with $\rm F$ or $\rm O$ atoms (red) \textit{e.g.} in the pre-bifurcated silver lattice (critical point of the conformal field theory (CFT)) of $\rm Ag_2F$ or $\rm Ag_2NiO_2$ respectively.\cite{schreyer2002synthesis, argay1966redetermination} (b), (c) and (d) A schematic of the electron (spins), indicated by up or down arrows, occupying each $4d^{10}$ and $5s$ orbital in the 3 degenerate states of $\rm Ag$ atoms labelled as $\rm Ag_{+1/2}$, $\rm Ag_{-1/2}$ and $\rm Ag_0$, whereby the core and valence electrons are shown as blue and green arrows respectively. Neglecting crystal field splitting, all $4d^{10}$ orbitals would have an equal probability to $sd$-hybridise with the $5s$ orbital, which serves as an arena for other interesting mechanisms. The energy gap between the $4d_z$ and $5s$ orbitals has been taken to be $\sim 3.5$ eV.\cite{blades2017evolution}}}\label{Fig_5}
\end{figure*}

\red{A typical crystal field splitting of $d$ orbitals in the prismatic environment\cite{huisman1971trigonal} is shown in \textbf{Figure 5(a)}, whereby in our case assuming completely filled $4d^{10}$ orbitals, $4d_{z^2}^2$ has the lowest energy, 
followed by degenerate $4d_{x^2 - y^2}^2$ and $4d_{xy}^2$ orbitals and finally degenerate $4d_{yz}^2$ and $4d_{xz}^2$ orbitals with the highest energy, as shown in \textbf{Figure 5(b)}. After hybridisation, the rest of the $4d$ orbitals are fully occupied with their energy levels unaltered, except for the newly formed $\rm Ag$ degenerate states, $4d_{z^2}^25s^1$ ($\rm Ag_{+1/2}$) and $4d_{z^2}^15s^2$ ($\rm Ag_0, Ag_{-1/2}$) given in \textbf{Figure 5(c)} and \textbf{Figure 5(d)}, which are only distinguishable via their valency states. Similar considerations apply in the case of linear/dumbbell coordination \textit{etc}. Thus, neglecting crystal field splitting, all $4d^{10}$ orbitals must have 
near-equal probability to $sd$-hybridise with the $5s$ orbital, which serves as an arena for other interesting mechanisms.} Nonetheless, whilst this crystal field splitting is not a requisite for $sd$-hybridisation, such additional mechanisms on the honeycomb lattice require the isolated $4d_{z^2}$ orbital to play the role of $2p_z$ orbital in graphene, whereby the crystal field splitting of the $4d$ orbitals together with $sd$ hybridisation play a role analogous to $sp^2$ hybridisation.\cite{allen2010honeycomb} Consequently, most properties such as pseudo-spin, pseudo-magnetic field \textit{etc.} exhibited by the itinerant $p_z^1$ electrons on graphene\cite{allen2010honeycomb, mecklenburg2011spin, georgi2017tuning, kvashnin2014phase} can be mapped to the electron properties on the $4d_{z^2}^1$ electrons in $\rm Ag$.\cite{kanyolo2022conformal} 

Meanwhile, $sd$-hybridisation guarantees that the electron properties of all the $4d_{z^2}5s$ (valence) electrons can be associated with the $\rm Ag$ atoms themselves -- a feature not 
expected in graphene-based systems. In other words, the theory of such electrons in graphene is $1 + 2$ dimensional quantum electrodynamics in a static space-time background whereas the dynamical nature of the silver atoms (which can diffuse during intercalation/de-intercalation processes) in contrast to carbon atoms in graphene lead instead to a dynamical space-time (quantum gravity), whereby a bifurcated honeycomb lattice introduces $1 + 3$ dimensions (same as Einstein gravity).\cite{kanyolo2022conformal} Due to the complexity involved in the description of dynamical space-times, we shall consider a static and flat space-time background. Proceeding, we can summarise the oxidation states to be searched for experimentally by $\rm Ag^{1+}/Ag_{+1/2}$, $\rm Ag^{1-}/\rm Ag_{-1/2}$ and $\rm Ag^{2+}/Ag_0$, where $\rm Ag_0$ is also given by $4d^95s^2$, reflecting its predisposition to form $\rm Ag^{2+}$ oxidation state instead of $\rm Ag^{1-}$. \red{Indeed, $\rm Ag^{2+}$ has been experimentally observed in $\rm Ag^{2+}F_2^{1-}$ and its $4d_{z^2}^15s^0$ character verified by 
its anti-ferromagnetism arising from $\rm Ag^{2+} - F_{(2)} - Ag^{2+}$ super-exchange interactions\cite{grzelak2017metal, kurzydlowski2021fluorides}, whilst $\rm Ag^{1-}$ has been reported in coinage metal cluster ions such as $\rm Ag^{1-}_N$, where $N = 1, 2, 3$ \textit{etc}.\cite{ho1990photoelectron, minamikawa2022electron, dixon1996photoelectron, schneider2005unusual}}

Nonetheless, we are interested in lifting these degeneracies by symmetry breaking. In nuclear physics, despite their different masses and charges, the proton and the neutron are essentially degenerate particles related by a degree of freedom that transforms appropriately under the SU($2$) gauge group known as iso-spin.\cite{yang1954conservation} This is also the case for leptons in the standard model of particle physics such as the electron and the neutrino, well described by SU($2$)$\times$U($1$) symmetry breaking, where U($1$) is the symmetry group also exhibited by Maxwell's equations.\cite{zee2010quantum, weinberg1967model} In our present case, in order to ensure the degenerate states of $\rm Ag$ are treated as fermions, we are interested in $\rm Ag_{+1/2}, Ag_{-1/2}$ and $\rm Ag^{2+}$ since they have an odd number of electrons in their orbitals. Note that, the actual fermionic or bosonic character of these states will differ by including the spin of the protons and neutrons in $\rm Ag$, reflecting the fact that we are only interested in the bonding properties of the degenerate states with dynamics inherited solely from the valence electrons. In our formalism, the degeneracy between $\rm Ag_{+1/2}$ and $\rm Ag_{-1/2}$ corresponds to isospin, which introduces the gauge group, SU($2$). Moreover, the degeneracy between $\rm Ag^{2+}$ and $\rm Ag_{-1/2}$ can be treated on the honeycomb lattice as the spin degree of freedom known in graphene physics as pseudo-spin.\cite{mecklenburg2011spin, georgi2017tuning} To yield results consistent with experimental observations in bilayered frameworks, $\rm Ag_{+1/2}$ and $\rm Ag_{-1/2}$ have left-handed chirality whereas $\rm Ag^{2+}$ is right-handed.
 
By promoting/demoting an electron 
using a neutral gauge field, $\rm Ag_{+1/2}$ can transform into $\rm Ag_{-1/2}$ ($4d^{10}5s^1 \rightarrow 4d^95s^2$) and \textit {vice versa} but never into $\rm Ag^{2+}$, since this is an oxidation state requiring the loss of electrons. We shall treat the entangled state of the electron pair and the neutral gauge field as a charged $W^{\mu}_{\pm} = \frac{1}{\sqrt{2}}(W_1^{\mu} \pm iW_2^{\mu})$ gauge boson, responsible for this transition, where $W^{\mu} = W^{\mu}_a\tau_a$ is a gauge field transforming under SU($2$) ($\tau_a = \sigma_a/2$, $\sigma_a = (\sigma_1, \sigma_2, \sigma_3)$ are the Pauli matrices). This transition has to be of the order $\sim 3.5$ eV, corresponding to the mass of $W_{\pm}^{\mu}$. We shall also consider a screened and an un-screened Coulomb potential, $Z_{\mu}$ and $p_{\mu}$ respectively obtained by the mixing,
\begin{align}\label{Weinberg_eq}
    \begin{pmatrix}
A^{\mu}\\ W_3^{\mu}
\end{pmatrix}
= 
\frac{1}{\sqrt{q_{\rm e}^2 + q_{\rm w}^2}}
\begin{pmatrix}
q_{\rm w} & -q_{\rm e}\\ 
q_{\rm e} & q_{\rm w}
\end{pmatrix}
\begin{pmatrix}
p^{\mu}\\ Z^{\mu}
\end{pmatrix},
\end{align}
where $q_{\rm e} \simeq 1.602 \times 10^{-19}$ C is the elementary (U($1$)) charge, $A_{\mu}$ is the electromagnetic field, $q_{\rm w}$ is the SU($2$) charge and,
\begin{align}\label{q_eff_eq}
    q_{\rm eff} = \frac{q_{\rm w}q_{\rm e}}{\sqrt{q_{\rm w}^2 + q_{\rm e}^2}},
\end{align}
is the effective coupling/charge to $p_{\mu}$. Now, introducing the charge operator $Y$ for the U($1$) electric charge the generator of the effective charge, $Q$ for the field $p_{\mu}$ is given by (a variant of) the Gell-Mann-Nishijima relation\cite{zee2010quantum}, 
\begin{align}
    Q = 2I + Y,
\end{align}
with $I = \tau_3 = \sigma_3/2 = \pm 1/2, 0$ the emergent iso-spins of left-handed ($\rm Ag_{+1/2} (\textit {I} = +1/2)$, $\rm Ag_{-1/2} (\textit {I} = -1/2)$) and right-handed ($\rm Ag^{2+} (\textit {I} = 0)$) fermions respectively corresponding to half the valency/oxidation state of the left-handed Ag cations, $\rm Ag_{+1/2} \rightarrow Ag^{1+}, Ag_{-1/2} \rightarrow Ag^{1-}$ but vanishes for the right-handed $\rm Ag^{2+}$ oxidation state. Meanwhile, since $W^{\mu}_{\pm}$ carries the iso-spin by transforming $\rm Ag_{+1/2}$ into $\rm Ag_{-1/2}$ and \textit {vice versa}, it has iso-spin, $I = \pm 1$, with an effective charge of $Q = \pm 1$ and electric charge, $Y = 0$. Thus, the effective charge of $W^{\mu}_{\pm}$ is a result of the entangled state with the valence electron, as earlier remarked. Moreover, charge and iso-spin conservation for the interaction between $W_{\pm}^{\mu}$ and the left-handed $\rm Ag$ states yields for $\rm Ag_{+1/2}$, $Y = 0$, $Q = +1$ and for $\rm Ag_{-1/2}$, $Y = 0$, $Q = -1$.  

\begin{table*}[!t]
\caption{\textbf{Table I}. Comparison between the various charges and spins of the relevant cations and fields in the theory of SU($2$)$\times$ U($1$) spontaneous symmetry breaking in $\rm Ag$ bilayered materials.}\label{Table_1}
    \centering
    \resizebox{\textwidth}{!}{
    \begin{tabular}{||c||c||c|c|c||c||}
    \hline\hline
    \textbf{cation, field $\diagdown$ spin, charge} & spin & iso-spin & U($1$) charge & effective charge & pseudo-spin\\
    & ($s$) & ($I$) & ($Y$) & ($Q = 2I + Y$) & ($s'$)\\
    \hline
    $\rm Ag^{1+}/Ag_{+1/2}$ & $\pm 1/2$ & $+1/2$ & $0$ & $+1$ & $0$\\
    $\rm Ag^{1-}/Ag_{-1/2}$ & $\pm 1/2$ & $-1/2$ & $0$ & $-1$ & $\mp 1/2$\\
    $\rm Ag^{2+}/Ag_0$ & $\pm 1/2$ & $0$ & $+2$ & $+2$ & $\pm 1/2$\\
    $p^{\mu}$ & $\pm 1$ & $0$ & $0$ & $0$ & $0$\\
    $W_{\pm}^{\mu}$ & $\pm 1$ & $\pm 1$ & 0 & $\pm 1$ & $0$\\
    $Z^{\mu}$ & $\pm 1$ & $0$ & $0$ & $0$ & $0$\\
    $\Psi, \Psi^*$ & $0$ & $-1/2, +1/2$ & $-2, +2$ & $-3, +3$ & $\mp 1, \pm 1$\\
    \hline\hline
    \end{tabular}}
\end{table*}

A summary of these charges and spins for the cations and the gauge bosons has been availed in \textbf{Table I}. The appropriate Lagrangian is given by,
\begin{multline}
    \mathcal{L} = \int d^{\,3}x\, \frac{1}{4}\left(\left |\partial_{\mu}\vec{W}_{\nu} + q_{\rm w}(\vec{W}_{\mu}\times\vec{W}_{\nu}) \right |^2 + \left(\partial_{\mu}A_{\mu} - \partial_{\nu}A_{\mu}\right)^2 \right)\\
    + \int d^{\,d}x\,\left (\overline{\psi}_{\rm L}\gamma^{\mu}(i\partial_{\mu} + q_{\rm w}W_{\mu} + \frac{Y}{2}q_{\rm e}A_{\mu})\psi_{\rm L} + \overline{\psi}_{\rm R}\gamma^{\mu}(i\partial_{\mu} + \frac{Y}{2}q_{\rm e}A_{\mu})\psi_{\rm R}\right)\\
    + \int d^{\,d}x\,\left(\frac{1}{2}\left |\left (\partial_{\mu} + iq_{\rm w}W_{\mu} + \frac{Y}{2}iq_{\rm e}A_{\mu}\right )\phi \right|^2 - \alpha\left(v^2 - \frac{1}{4}(\phi^{\dagger}\phi)\right)^2 + \gamma\overline{\psi}_{\rm R}\phi^{\dagger}\psi_{\rm L} + h.c.\right ),
\end{multline}
where $\psi_{\rm L}^{\rm T} = \rm (Ag_{+1/2}, Ag_{-1/2})$, $\psi_{\rm R}^{\rm T} = \rm Ag^{2+}$, $\overline{\psi}_{\rm L/R} = \psi_{\rm L/R}^{*\rm T}\gamma^0$, $\phi^{\dagger} = (0, \Psi^*|T_{\rm c}|^{-1/2})$, the superscript $\rm T$ is the transpose, $h.c.$ stands for Hermitian conjugate (of $\gamma\overline{\psi}_{\rm R}\phi^{\dagger}\psi_{\rm L}$), $\gamma^{\mu}$ are the gamma matrices, $\alpha$, $v$, $\gamma$ are 
constants and $|\Psi|^2$ is a $d$D condensate satisfying\cite{kanyolo2022conformal},
\begin{align}
    \int d^{\,d}x\,|\Psi|^2 = 2k,\,\, \langle \langle k \rangle \rangle + \Delta(d) = 0,
\end{align}
where $k \in \mathbb{N}$ is the number of cationic vacancies\cite{kanyolo2020idealised, kanyolo2021honeycomb, kanyolo2022cationic} behaving like a condensate, $\Delta(d) = (d - 2)/2$ is the scaling dimension of a mass-less scalar conformal field theory\cite{francesco2012conformal, kanyolo2022conformal} and $d = 2, 3$ with a suitable average $\langle\langle \cdots \rangle\rangle$ corresponding to a normalised Mellin transform\cite{kanyolo2022conformal} of the thermal average $\langle k \rangle = \langle a^{\dagger}a \rangle$ (Here, $a^{\dagger}, a$ are the quantum harmonic oscillator raising, lowering operators).

The desired field theory results can be computed in parallel to electroweak theory (specifically, the 
lepton interactions in the standard model\cite{zee2010quantum, weinberg1967model}, with the exception of $Y = 0$ for the left-handed cations, as in \textbf{Table I}, instead of $Y = -1$ for the leptons in standard model, which leaves $\rm Ag^{1+}$ charged ($Q \neq 0$) unlike the neutrino ($Q = 0$)) \textit{albeit} in $1 + d$ dimensions ($d = 2, 3$) with the electromagnetic field, $A_{\mu}$ playing the role of the hyper-charge.\cite{zee2010quantum} Focusing only on the important features, the Lagrangian introduces a mass for $W_{\pm}^{\mu}$ and $Z^{\mu}$ ($m_W$ and $m_Z$ respectively) related by, 
\begin{align}\label{mass_WZ_eq}
    3.5 \, {\rm eV} \sim m_W = \frac{q_{\rm w}}{\sqrt{q_{\rm e}^2 + q_{\rm w}^2}}m_Z,
\end{align}
for $|\Psi| \propto \Delta(d) \neq 0$. Recall that $1/m_Z$ is the screening length of electromagnetic interactions within $\rm Ag$ bilayers. Thus, to determine $q_{\rm w}$, one can use eq. (\ref{q_eff_eq}), where $q_{\rm eff}$ is the effective charge of the cations as measured in experiments with bilayers (which is expected to differ from $q_{\rm e} = 1.602 \times 10^{-19}$ C), then proceed to solve for $q_{\rm w}$, which yields the mass $m_Z$ using eq. (\ref{mass_WZ_eq}). Moreover, only $\psi^{\rm T} = (\rm Ag^{2+}, Ag_{-1/2})$ acquires a mass term (potential/bonding energy),
\begin{align}\label{potential_eq}
    U(\Delta, T_{\rm c}) = \mathcal{L}_{\rm mass}(\overline{\psi}, \psi)  = \frac{\gamma}{|T_{\rm c}|^{1/2}}\int d^{\,d}x\,|\Psi|\overline{\psi}\psi \sim \frac{2\Delta(d)}{|T_{\rm c}|^{\Delta^*(d)}}\int d^{\,d}x\,(T - T_{\rm c})\overline{\psi}\psi \leq 0,
\end{align}
which we shall interpret as the origin of the argentophilic interaction between $\rm Ag^{2+}$ and $\rm Ag^{1-}$, where $\Delta^*(d) = (1 - 2\Delta(d))/2$, we have made the replacement, $\Psi^* \rightarrow |\Psi|\exp(-iS)$ with $S = 0$ in $\phi^{\dagger}$ for simplicity and we have introduced the critical exponent,
\begin{align}
   |\Psi(T, d)| = 2v(T, d)|T_{\rm c}|^{1/2} \sim 2\Delta(d)|T_{\rm c}|^{\Delta(d)}(T - T_{\rm c})/\gamma \geq 0,
\end{align}
with $v(T, d) = v \geq 0$ the constant appearing in the Lagrangian (related to $|\Psi|$ by minimising the Mexican hat potential with respect to $\phi, \phi^{\dagger}$), $T$ the temperature and $T_{\rm c}$ the critical temperature (acquired mass of $\psi$ in $d = 3$).\cite{domb2000phase} Note that, the exponent $\Delta(d)$ appearing in $|T_{\rm c}|$ is justified by dimensional analysis. Moreover, the mass of the fields $\phi, \phi^{\dagger}$ corresponds to $m_{\phi} = |\alpha|v^2$. Consequently, all mass terms vanish in $d = 2$ dimensions due to the scaling dimension, $\Delta(d = 2) = 0$, but are finite for $d = 3$ provided $T > T_{\rm c}$, since $\Delta(d = 3) = 1/2$. This represents a monolayer-bilayer phase transition for $T > T_{\rm c}$, where $T(\vec{x})$ is a temperature field that behaves like the Higgs field.\cite{zee2010quantum} Moreover, its bosons can be interpreted as phonons within the material arising from high temperature dynamics.  

The triumph herein is that the formalism satisfies (i), (ii) and (iii) above. Specifically, (ii) is satisfied for instance by writing\cite{kanyolo2022advances}, $\rm Ag_2^{1/2+}Ni^{3+}O_2^{2-} = Ag^{2+}Ag^{1-}Ni^{3+}O_2^{2-}$. Other sub-valent states such as $\rm Ag_3^{2/3+}Ni_2^{3+}O_4^{2-}$ represent a saturation or hybrid effect by the mass-less $\rm Ag^{1+}$ fermion\cite{kanyolo2022advances},
\begin{subequations}
\begin{align}
    \rm Ag^{2+}Ag^{1-}Ni^{3+}O_2^{2-} + Ag^{1+}Ni^{3+}O_2^{2-} \rightarrow  Ag^{2+}Ag^{1-}Ag^{1+}Ni_2^{3+}O_4^{2-} = Ag_3^{2/3+}Ni_2^{3+}O_4^{2-},
\end{align}
Lastly, $\rm Ag^{1+}Ni^{3+}O_2^{2-}$ cannot be bilayered since $\rm Ag^{1+}$ is mass-less in the theory. In the case of the present material in this study, assuming $\rm Te^{6+}$, the under-saturated bilayered material is expected to be given by ${\rm Ag^{2+}_2Ag^{1-}_2Ni}_2^{2+}{\rm Te}^{6+}{\rm O_6^{2-}} = \rm Ag_4^{1/2+}Ni_2^{2+}Te^{6+}O_6^{2-}$ or ${\rm Ag^{2+}_4Ag^{1-}_4Ni}_2^{2+}{\rm Te}^{4+}{\rm O_6^{2-}} = \rm Ag_8^{1/2+}Ni_2^{2+}Te^{4+}O_6^{2-}$ ($\rm Ag$ coordination to $\rm O$ is assumed prismatic) with Ag sub-valency $+1/2$, consistent with our XPS and XAS experimental observations ($\rm Ni^{2+}, Te^{4+}, Te^{6+}$). However, the STEM-EDX spectra results require the chemical formula, $\rm Ag_6Ni_2TeO_6$ for the $\rm Ag$-rich domain bilayered domains of the present material. Moreover, we can have, ${\rm Ag_4^{1+}Ag_2^{1-}Ni}_2^{2+}{\rm Te}^{6+}{\rm O_6^{2-}} = \rm Ag_6^{1/3+}Ni_2^{2+}Te^{6+}O_6^{2-}$ or ${\rm Ag_2^{2+}Ag_2^{1-}Ag_2^{1+}Ni}_2^{2+}{\rm Te}^{4+}{\rm O_6^{2-}} = \rm Ag_6^{2/3+}Ni_2^{2+}Te^{4+}O_6^{2-}$, where the former can be excluded theoretically from the bilayered frameworks since it lacks the right-handed, $\rm Ag^{2+}$ valency state and hence lacks the mass term responsible for the bilayered structure. Whilst not a requisite in the theory, we note that none of these materials contain both $\rm Te^{4+}$ and $\rm Te^{6+}$ as observed experimentally. To remedy this, we shall consider the hybrid, 
\begin{multline}
    {\rm Ag^{2+}_2Ag^{1-}_2Ni}_2^{2+}{\rm Te}^{6+}{\rm O_6^{2-} + Ag^{2+}_4Ag^{1-}_4Ni}_2^{2+}{\rm Te}^{4+}{\rm O_6^{2-}} \rightarrow\\
    {\rm Ag_6^{2+}Ag_6^{1-}Ni_4^{2+}Te^{4+}Te^{6+}O_{12}^{2-} = 2Ag_6^{1/2+}Ni_2^{2+}Te^{5+}O_6^{2-}},
\end{multline}
\end{subequations}
which is also bilayered, consistent with the theory. Ideally, the oxidation states $\rm Ag^{2+}$, $\rm Ag^{1-}$ and $\rm Ag^{1+}$ alongside the sub-valency states of $\rm Ag^{1/2+}$ or $\rm Ag^{2/3+}$, when present, should be observable in XPS or XAS data, provided distinguishability issues encountered such as the 
crystalline homogeneity (phase purity) of the as-prepared material encountered in the present work can be adequately addressed. Moreover, the energy gap of $\sim$ 3.5 eV should be existent in measurement data (XPS or XAS) whenever $\rm Ag^{1+}/Ag_{+1/2}$ is available and transmutes into $\rm Ag^{1-}/Ag_{-1/2}$ or \textit {vice-versa}. This should occur, for instance, in processes where a bilayered structure disintegrates into a monolayered structure (or \textit {vice versa}) by the emission or absorption of photo-electrons of order energy gap between the $5s$ and $4d_z$ orbitals, taken to be $\sim 3.5$ eV. Presently, we have neither tested nor observed such a monolayer-bilayer phase transition, which falls beyond the technical abilities brought to bear in the present work. For brevity, other considerations for additional sub-valent states not fully consistent with our observations in this paper will be considered elsewhere.\cite{kanyolo2022conformal, kanyolo2022advances}

Finally, topological aspects consistent with idealised model\cite{kanyolo2020idealised, kanyolo2021honeycomb, kanyolo2022cationic} can be calculated in $d = 2$ dimensions by\cite{zee2010quantum}, 
\begin{subequations}
\begin{align}
    \int \mathcal{D}[\overline{\psi},\psi]\exp(i\int dt\,\mathcal{L}) = \exp(iS')\exp(iS_{\rm CS}),\\
    S_{\rm CS} = \frac{q_{\rm eff}^2}{2}\int dt\int d^{\,2}x\,\left(\frac{1}{4\pi}\frac{T_{\rm c}}{|T_{\rm c}|}\varepsilon^{\mu\nu\sigma}p_{\mu}\partial_{\nu}p_{\sigma} + \phi^{\dagger}\phi\,p_{\mu}p^{\mu}\right),
\end{align}
\end{subequations}
where $\varepsilon^{\mu\nu\sigma}$ is the Levi-Civita symbol normalised as $\varepsilon^{012} = 1$ and we have kept leading terms with $p_{\mu}$ in $S_{\rm CS}$ (Chern-Simons action\cite{zee2010quantum, kanyolo2019berry}) and the rest in $S'$. The equations of motion yield, 
\begin{align}\label{CS_eq}
    \frac{1}{4\pi}\varepsilon^{\mu\nu\sigma}\partial_{\nu}p_{\sigma} = T_{\rm c}^{-1}|\Psi|^2p^{\mu},
\end{align}
where we have used $\phi^{\dagger}\phi = |\Psi|^2/|T_{\rm c}|$. Moreover, we can rescale $p_{\mu} \rightarrow p_{\mu}\sqrt{2}/q_{\rm eff}$ and set $\vec{p} = \vec{n}\times\vec{\nabla}\Phi$ where $p^{\mu} = (T_{\rm c}, \vec{p})$, $\vec{n} = (0, 0, 1)$ is the unit normal vector to the cationic honeycomb lattice and $\Phi$ is the Liouville field satisfying, $\nabla^2\Phi = -K\exp(2\Phi)$ with $K$ the Gaussian curvature.\cite{kanyolo2020idealised, kanyolo2022cationic} Thus, the time component of eq. (\ref{CS_eq}) corresponds to, $\vec{\nabla}\times\vec{p} = \nabla^2\Phi = -K\exp(2\Phi) = |\Psi|^2$, which satisfies Poincar\'{e}-Hopf theorem.\cite{kanyolo2022cationic} 
There are two geometries of the $\rm Ag$ lattice representing the honeycomb lattice that satisfy the no vacancy condition, $\int K\exp(2\Phi) = 0$: The flat torus ($K = 0$) and the two-torus $\int_{\mathbb{T}^2}K\exp(2\Phi) = 2\pi(2 - 2g) = -4\pi k = 0$ ($g$ corresponds to the genus of an emergent $2$D surface without boundary) with $K \neq 0$, related to each other by the transformation given in \textbf {Figure 4d}. This means that the critical point is two-fold degenerate. Coordinate transformations in this state correspond to conformal invariance (modular invariance\cite{kanyolo2022cationic}), which is broken for $g \neq 1$. This can be achieved either by the creation of vacancies $g > 1$ or by the system lowering its ground state energy by exploiting the additional $g = 0$ state. This additional state is the energetically more stable configuration away from the critical point ($T > T_{\rm c}$), \textit{albeit} inaccessible in $d = 2$ dimensions. In other words, the bipartite honeycomb lattice lifts this degeneracy by bifurcation - interpreted as a variant of the theorem of Peierls (1D) or Jahn-Teller (3D).\cite{garcia1992dimerization, jahn1937stability, kanyolo2022conformal} 

Moreover, the opposite convention for the temperature in the phase transition (transition happens for $T > T_{\rm c}$ instead of the conventional $T < T_{\rm c}$) is indicative of the emergence of a 3D gravitational description of the potential in eq. (\ref{potential_eq}), 
\begin{align}
    U(\Delta, T_{\rm c}) \simeq -A(\Delta, T_{\rm c})/R^2 + B(\Delta, T_{\rm c})/R^3,
\end{align}
where $A(\Delta, T_{\rm c})$, $B(\Delta, T_{\rm c})$ are constants independent of $R$ and $R_{\rm disp.} = 3B/2A \simeq 2.8$ \AA\,\, is the observed displacement due to the bifurcation satisfying the condition, $\partial U/\partial R|_{R = R_{\rm disp.}} = 0$, which scales as Newtonian acceleration/gravity.\cite{kanyolo2022conformal} This can be interpreted as a dual description of the Liouville conformal field theory (CFT) vacancy states ($k > 1$) in the spirit of gravity-CFT duality.\cite{kanyolo2022conformal} Finally, this potential corresponds to a Hamiltonian of the honeycomb lattice pseudo-spin degrees of freedom (a 1D Ising model of pseudo-spins interacting with a pseudo-magnetic field corresponding to the Gaussian curvature ($g = 0$) and the Heisenberg coupling taken to be the Ruderman–Kittel–Kasuya–Yosida (RKKY) interaction).\cite{kanyolo2022cationic} Further discussion on the conformal field theoretic nature of the monolayer-bilayer phase transition has been availed in 
\textbf{Supplementary Note 2}.

\section{\label{section: Conclusion} Conclusion}

We report the topochemical synthesis at 250 $^{\circ}$C (and under ambient pressure) of a new honeycomb layered oxide with a global average composition of ${\rm Ag_2}M_2{\rm TeO_6}$ ($M = \rm Ni, Mg, \textit{etc}$), manifesting Ag-rich and Ag-deficient domains. Aberration-corrected transmission electron microscopy reveals the Ag-rich crystalline domains with a composition of ${\rm Ag_6}M_2{\rm TeO_6}$ to exhibit Ag atom bilayers with aperiodic stacking disorders. X-ray absorption spectroscopy and X-ray photoelectron spectroscopy ascertain sub-valent Ag states innate in the bilayer ${\rm Ag_6}M_2{\rm TeO_6}$ domains with the origin rationalised to arise from spontaneous symmetry breaking of degenerate mass-less chiral fermion states of silver. Such a phase transition yields three oxidation states of silver ($\rm Ag^{1+}$, $\rm Ag^{1-}$ and $\rm Ag^{2+}$) on the honeycomb lattice, based on the occupancy of their $4d_{z^2}$ and $5s$ orbitals, and a mass term between $\rm Ag^{2+}$ and $\rm Ag^{1-}$ responsible for the bilayered structure. From this theoretical point of view, we acquire an intuitive picture for the origin of the argentophilic bond between $\rm Ag$ pairs responsible for stabilising the bilayers in $\rm Ag$-based materials with matching characteristics.\cite{johannes2007formation, yoshida2020static, taniguchi2020butterfly, yoshida2011novel, matsuda2012partially, yoshida2008unique, yoshida2006spin, beesk1981x} Moreover, since the 3.5 eV is the energy cost of $sd$-hybridisation which leads to degeneracy of the valence band ($5s$-orbitals) and conduction bands ($4d_{z^2}$-orbitals) on the honeycomb lattice, lifting this degeneracy corresponds to a metal-semiconductor/metal-insulator phase transition, in the spirit of Peierls instability\cite{garcia1992dimerization} (analogous to Cooper-pair instability, which is related to paired electrons), resulting in an energy gap.\cite{little1964possibility} Whilst this energy gap will differ from the 3.5 eV depending on the nature of the instability, in the case of $\rm Ag$-bilayered materials, the bilayered structure 
ought to be able to disintegrate into a monolayered structure and \textit {vice versa} by the emission or absorption of photoelectrons of order energy gap between the $4d_{z^2}$ and $5s$ orbitals, taken to be $\sim 3.5$ eV.\cite{blades2017evolution} Presently, we have neither tested nor observed such a monolayer-bilayer (conductor-semiconductor/insulator) phase transition, or related effects which falls beyond the technical abilities brought to bear in the present work. 

Nonetheless, the achieved experimental and theoretical insights not only promise to augment the literature space of $\rm Ag$-based honeycomb layered oxides structures, mechanisms and functionalities, but also are poised to inspire innovative applications for these next-generation functional materials. Ultimately, we regard the silver-based honeycomb layered tellurate as a pedagogical platform for further inquiry into the role of geometric features and non-commutative electromagnetic interactions, which go beyond energy storage applications.

\section{\label{section: Methods} Methods}

\subsection{Synthesis of materials}

Honeycomb layered oxides embodying the global composition of ${\rm Ag_2}M_2{\rm TeO_6}$ (where $M = \rm Ni, Co, Cu, Zn, Mg$ and $\rm Ni_{0.5}Co_{0.5}$) were synthesised via the topochemical ion-exchange of ${\rm Na_2}M_2{\rm TeO_6}$ precursors alongside a molten flux of $\rm AgNO_3$ at 250 $^{\circ}$C for 99 hours in air, based on the following reaction:
\begin{align}
    \,{\rm Na_2{\it M}_2TeO_6} ({\it M}= \rm Ni, Mg, Co, Cu, Zn) + 2\,{\rm AgNO_3} \rightarrow {\rm Ag_2{\it M}_2TeO_6} + 2\,{\rm NaNO_3}.
\end{align}
A 4-fold molar excess amount of $\rm AgNO_3$ was used to facilitate a complete ion-exchange reaction. To dissolve the residual nitrates ({\it i.e.}, $\rm NaNO_3$ byproduct and remaining $\rm AgNO_3$), the resulting product was thoroughly washed with distilled water, vigorously stirred with a magnetic mixer and thereafter filtered and dried. The resulting ${\rm Ag_2}M_2{\rm TeO_6}$ displayed varied colours distinct from the precursor materials confirming the completion of the ion-exchange reaction (see \textbf{Supplementary Figure 32}). Note that, ${\rm K}_2M_2\rm TeO_6$ or ${\rm NaK}M_2\rm TeO_6$ can also be used as precursor materials. However, due to the inherent hygroscopicity of potassium-based honeycomb layered compounds, ${\rm Na_2}M_2{\rm TeO_6}$ was selected as the precursors for a facile and scalable $\rm Na^+/Ag^+$ ion-exchange reaction. ${\rm Na_2}M_2{\rm TeO_6}$ precursor materials were prepared using the high-temperature solid-state reaction route detailed in literature.\cite{evstigneeva2011new, sankar2014crystal, chen2020high, masese2021unveiling, bera2020temperature}

\subsection{X-ray diffraction (XRD) analyses}

Conventional XRD (CXRD) measurements were conducted using a Bruker D8 ADVANCE diffractometer to ascertain the crystallinity of the as-prepared ${\rm Ag_2}M_2{\rm TeO_6}$ powder samples. Measurements were performed in Bragg-Brentano geometry mode with Cu-{\it K}$\alpha$ radiation. Synchrotron XRD (SXRD) measurements were performed to acquire high-resolution data of as-prepared ${\rm Ag_2}M_2{\rm TeO_6}$. SXRD experiments were performed at BL8S2 of Aichi SR Center. XRD \textit {ex situ} measurements of pristine and charged ${\rm Ag}_{2-x}\rm Ni_2TeO_6$ and ${\rm Ag}_{2-x}\rm NiCoTeO_6$ electrodes were collected in Bragg-Brentano geometry using a Cu-{\it K}$\alpha$ monochromator. Prior to performing XRD measurements, the electrodes were thoroughly washed using super-dehydrated acetonitrile and subsequently dried in an argon-purged glove box.

\subsection{Morphological and physicochemical characterisation}

Field emission scanning electron microscope (JSM-7900F) was used to analyse the morphologies of the obtained powder samples. Energy dispersive X-ray (EDX) imaging technique was used to assess the constituent elements of the obtained powders. Quantitative assessment of the chemical compositions was performed using inductively coupled plasma absorption electron spectroscopy (ICP-AES). Further, the density of the as-prepared powders was determined via pycnometric measurements (see \textbf {Supplementary Table 2}). The information obtained was used to calculate the pellet compactness prior to the electrochemical impedance spectroscopic measurements.
Specimens for atomic-resolution transmission electron microscopy (TEM) were prepared by an Ar-ion milling method using a GATAN PIPS (Model 691) precision ion-milling machine after embedding them in epoxy glue under an Ar atmosphere. High-resolution scanning TEM (STEM) imaging and electron diffraction patterns were obtained using a JEOL JEM-ARM200F incorporated with a CEOS CESCOR STEM Cs corrector (spherical aberration corrector). The acceleration voltage was set at 200 kV. Electron microscopy measurements were conducted along various zone axes (namely, [100], [010] and [310] zone axes). To mitigate beam damage to the samples, a low electron-beam dosage (STEM probe current value of 23 pA) was used with short-exposure times. The probe-forming convergence angle was 22 mrad. High-angle annular dark-field (HAADF) and annular bright-field (ABF) STEM snapshots were taken simultaneously at nominal collection angles of 90$\sim$370 mrad and 11$\sim$23 mrad, respectively. To reduce the possibility of image distortion induced by the specimen drift during the scan, a quick sequential acquisition technique was conducted to observe the atomic structures. It is important to mention here that images of the $\rm Ag_2Zn_2TeO_6$ could not be obtained on account of its low crystallinity (amorphous nature), as was further affirmed by XRD measurements. Attempts to improve crystallinity of the sample by annealing at temperatures below the decomposition regime proved elusive. For accurate localisation of metal atoms in the obtained STEM maps, about 20 STEM images were recorded sequentially with an acquisition time of about 0.5 s per image, after which the images were aligned and superimposed into one image. STEM-EDX (energy-dispersive X-ray spectroscopy) spectrum images were obtained with two JEOL JED 2300T SDD-type detectors with 100 $\rm mm^{2}$ detecting area whose total detection solid angle was 1.6 sr. Elemental maps were extracted using Thermo Fisher Scientific Noran (NSS) X-ray analyser. Reproducibility measurements on various crystallites were also performed using TITAN cubed G2 60-300 TEM (FEI Company) (acceleration voltage: 300 kV) coupled with an EDX, in which the EDX measurements were conducted by using Super-X (Bruker).

\subsection{Electrochemical measurements}

Fabrication of the composite electrode was performed by mixing the as-prepared $\rm Ag_2Ni_2TeO_6$ and $\rm Ag_2NiCoTeO_6$ polycrystalline powders with polyvinylidene fluoride (PVdF) binder and acetylene black (conductive carbon) at a weight ratio of 70:15:15. The mixture was suspended in $N$-methyl-2-pyrrolidinone (NMP) to attain viscous slurry samples, which were then cast on aluminium foil with a typical mass loading of $\sim$5 mg cm$^{-2}$, before drying under vacuum. Electrochemical measurements were assessed using CR2032-type coin cells using $\rm Ag_2Ni_2TeO_6$ and $\rm Ag_2NiCoTeO_6$ composite electrodes as the cathodes (working electrodes) in $\rm Ag$ half-cells and $\rm Li$ half-cells. Glass fibre discs were used as separators alongside electrolytes consisting of 0.1 mol dm$^{-3}$ silver bis(trifluoromethanesulphonyl)imide ($\rm Ag$TFSI) in 1-methyl-1-propylpyrrolidinium bis(trifluoromethanesulphonyl)imide ($\rm Pyr_{13}TFSI$) for the $\rm Ag$ half-cells and 0.5 mol dm$^{-3}$ lithium bis(trifluoromethanesulphonyl)imide (LiTFSI) in $\rm Pyr_{13}TFSI$ ionic liquid as electrolyte for the $\rm Li$ half-cells. The coin cells were assembled in an Ar-filled glove box (MIWA, MDB-1KP-0 type) with oxygen and water contents maintained below 1 ppm. All electrochemical measurements were performed at room temperature. Galvanostatic cycling was done at a current rate commensurate to C/10 (10 being the necessary hours to charge to full theoretical capacity). The upper cut-off voltage was set at 1.5 V for the $\rm Ag$ half-cells, or 4.8 V as for the $\rm Li$ half-cells.

\subsection{Thermal stability measurements}

A Bruker AXS 2020SA TG‐DTA instrument was used to perform thermogravimetric and differential thermal analysis (TG‐DTA). Measurements were performed at a ramp rate of 5 $^{\circ}$C $\rm min^{-1}$ under argon using a platinum crucible. Measurements were performed in the temperature ranges of 25 $^{\circ}$C to 900 $^{\circ}$C. 

\subsection{X-ray photoelectron spectroscopy (XPS) measurements}

XPS measurements were performed on pristine $\rm Ag_2Ni_2TeO_6$, $\rm Ag_2NiCoTeO_6$, charged ${\rm Ag}_{2-x}\rm Ni_2TeO_6$ and charged $\rm Ag_{2-{\it x}}NiCoTeO_6$ electrodes to ascertain the valency state upon silver-ion extraction. The electrodes were intimately washed with super-dehydrated acetonitrile and dried inside an argon-filled glove box, prior to undertaking XPS analyses at ${\rm Ag}\,\,3d$, ${\rm Co}\,\,2p$, ${\rm Te}\,\,3d$ and ${\rm Ni}\,\,2p$ binding energies. A hermetically sealed vessel was used to transfer the electrode samples into the XPS machine (JEOL(JPS-9030) equipped with both Mg {\it K}$\alpha$ and Al {\it K}$\alpha$ sources). For clarity, XPS analyses at ${\rm Te}\,\,3d$, ${\rm Ag}\,\,3d$ and ${\rm Co}\,\,2p$ binding energies were conducted using the Al {\it K}$\alpha$ source, whereas the Mg {\it K}$\alpha$ source was used for analyses at ${\rm Ni}\,\,2p$ binding energies. The electrodes were etched by an Ar-ion beam for 10 s to eliminate the passivation layer at the surface. The attained XPS spectra were fitted using Gaussian functions, and data processing protocols were performed using COMPRO software.

\subsection{X-ray absorption spectroscopy (XAS) measurements}

Charged ${\rm Ag}_{2-x}\rm Ni_2TeO_6$ electrodes were hermetically sealed in packets inside an Ar-purged glove box. The XAS spectra were measured in the $\rm Ni$ {\it K}-edge and Co {\it K}-edge energy region (at room temperature) in transmission mode at beamline BL14B2 of the SPring-8 (Japan) synchrotron facility. Athena package was used to treat the raw X-ray absorption data, as is customary. As for the O {\it K}-edge measurements, the (dis)charged electrode samples were transferred to a measurement vacuum chamber without air exposure. The spectra were measured in fluorescence yield mode (which is sensitive to the bulk state of a sample) using the beamline facility (BL1N2) of Aichi Synchrotron Radiation Center located at Aichi (Japan).

\subsection{Computational methods}

The charge density for $\rm Ag_2Ni_2TeO_6$, $\rm Ag_6Ni_2TeO_6$, $\rm AgCl$ and $\rm Ag_2O$ were optimised to be self-consistent (with a threshold of $10^{-7}$ eV) using the density functional theory (DFT) formalism with generalised gradient approximation (GGA), incorporating on-site Coulomb parameters and dispersion force correction. \red{The DFT calculations were performed by Vienna \textit {ab-initio} Simulation Package (VASP) programme.\cite{kresse1993ab, kresse1994ab, kresse1996efficient, kresse1996efficiency} The inner core region was assessed using the projector-augmented-wavefunction method.\cite{kresse1999ultrasoft} Thus, the Kohn-Sham equations\cite{kohn1965self} were solved only for the valence electrons.} The number and occupancy of (valence) electrons was set as follows: $\rm Ag$ ($4d^{10}5s^1$), $\rm Ni$ ($3d^84s^2$), $\rm Te$ ($5s^25p^4$), $\rm Cl$ ($3s^23p^5$) and $\rm O$ ($2s^22p^4$). Thus, other degenerate states of silver encountered in the present work were not considered in our preliminary simulation efforts 
availed in \textbf{Supplementary Information} (\textbf{Supplementary Figure 26} and \textbf {Supplementary Note 1}).

\subsection{Ionic conductivity measurements}

As-prepared $\rm Ag_2Mg_2TeO_6$ and $\rm Ag_2Ni_2TeO_6$ powder samples were uniaxially pressed into pellets with diameters of 10 mm under a pressure of about 40 MPa. Stainless steel (SUS) was used as the current collector. The pellet densities were approximately 74 \% and 84 \% of the theoretical ceramic densities, respectively. Electrochemical impedance measurements were done using a two-probe alternating current (a.c.) impedance spectroscopy (VSP-300 (Bio-Logic Science Instruments Corp.)) over a frequency ranging from 100 mHz to 3 MHz at a perturbation of 10 mV. Impedance spectroscopic data was initially collected at 25 $^{\circ}$C and thereafter between 30 $^{\circ}$C and 100 $^{\circ}$C, with impedance scans taken every 15 $^{\circ}$C. Ionic conductivities (of the bulk) recorded at various temperatures were obtained by Nyquist plot fittings. Nyquist plots displayed typical behaviour of ion-conducting material, which includes a semicircle at high frequencies and a linear spike at low frequencies. The activation energy ($E_{\rm a}$) for silver-ion conduction was calculated through a linear fitting of the bulk ionic conductivity values at various temperatures by incorporating the well-established Arrhenius equation, $\sigma = \sigma_0\exp(-E_{\rm a}/k_{\rm B}T)$ plotted in its log form (straight line equation: $\log(\sigma T) = -E_{\rm a}/k_{\rm B}T + \log(\sigma_0 T)$) versus inverse temperature, $1/T$ with a gradient $-E_{\rm a}/k_{\rm B}$ and $y$-intercept $\log(\sigma_0 T)$. Here, $\sigma$ denotes the temperature-contingent ionic conductivity, $\sigma_0$ as the absolute ionic conductivity (at zero temperature), $E_{\rm a}$ represents the activation energy (in this case, for silver-ion conduction), whilst $k_{\rm B}$ and $T$ are the Boltzmann constant and temperature respectively. All equivalent circuits of the Nyquist plots were fitted using the EC-Lab software package Z-fit.

\begin{addendum}
 
\item[Acknowledgments]

We thank Ms. Shinobu Wada and Mr. Hiroshi Kimura for the unrelenting support in undertaking this study. We gratefully acknowledge Ms. Kumi Shiokawa, Mr. Masahiro Hirata and Ms. Machiko Kakiuchi for their advice and technical help as we conducted the syntheses, electrochemical and XRD measurements. This work was supported by the TEPCO Memorial Foundation and AIST-Ritsumeikan University Fusion Seeds Sprout Program 2021. In addition, this work was also conducted in part under the auspices of the Japan Society for the Promotion of Science (JSPS KAKENHI Grant Numbers 19K15686, 20K15177 and 21K14730) and the National Institute of Advanced Industrial Science and Technology (AIST). T. Masese. and G. M. Kanyolo are grateful for the unwavering support from their family members (T. Masese.: Ishii Family, Sakaguchi Family and Masese Family; G. M. Kanyolo: Ngumbi Family). 

\item[Name Abbreviations] T.M, G.M.K., Y.M., M.I., N.T., J.R., S.T., K.T., Z.-D.H, A.A., H.U., K.K., K.Y., C.T., Y.O., H.K., T.S. 

\item[Author Contribution] T.M. and G.M.K. planned the project; T. Masese supervised all aspects of the research with help from G.M.K. and Z.-D.H; J.R. prepared the honeycomb layered oxide materials with the help from T.M.; N.T., Y.M., M.I. and T.S. acquired and analysed TEM data with input from H.S., G.M.K. and T.M.; K.Y. and T.M. performed the electrochemical measurements with input from G.M.K., A.A., H.S. and Z.-D.H; Y.O. acquired the high-resolution X-ray diffraction data; H.U., C.T. and H.K. helped in the analyses of the X-ray diffraction data; S.T. and Y.O. acquired X-ray absorption spectroscopy (XAS) data and conductivity data; K.Y. performed X-ray photoelectron spectroscopic measurements with input from T.M. and G.M.K; K.K. performed and analysed the thermal stability measurements; K.T. performed DFT calculations with input from G.M.K. and T.M.; The theoretical discussion including the mathematical framework for the origin of the bilayers in $\rm Ag$-based tellurates was entirely conceived and written by G.M.K. and T.M.; The manuscript was written by Y.M., J.R., T.S., K.T., G.M.K., Z.-D.H and T.M. All authors contributed to discussions and were given the chance to make comments and contributions pertaining the content in the manuscript and accompanied \textbf{Supplementary Information}.

\item[Data Availability]

Additional data that support the findings of this study are available on reasonable request from the corresponding authors.

\item[Competing Interests] 
The authors declare that they have no known competing financial interests or personal relationships that could have unethically impacted the rigour and scientific methods employed in this work.
 
\item[Correspondence] 
Correspondence and material requests should be addressed to:\\
Titus Masese (titus.masese@aist.go.jp) and Godwill Mbiti Kanyolo (gmkanyolo@mail.uec.jp).\\
Phone: +81-72-751-9224; Fax: +81-72-751-9609

\end{addendum}


\begin{thebibliography}{99}

\bibitem{kanyolo2021honeycomb}Kanyolo, G., Masese, T., Matsubara, N., Chen, C., Rizell, J., Huang, Z., Sassa, Y., M\aa nsson, M., Senoh, H. \& Matsumoto, H. Honeycomb layered oxides: structure, energy storage, transport, topology and relevant insights. {\em Chemical Society Reviews}. \textbf{50}, 3990-4030 (2021)

\bibitem{kanyolo2022advances}Kanyolo, G., Masese, T., Alshehabi, A. \& Huang, Z. Advances in honeycomb layered oxides: Syntheses and Characterisations of Pnictogen- and Chalcogen-Based Honeycomb Layered Oxides. {\em ArXiv Preprint ArXiv:2207.06499}. (2022)

\bibitem{house2020}House, R., Maitra, U., Perez-Osorio, M., Lozano, J., Jin, L., Somerville, J., Duda, L., Nag, A., Walters, A., Zhou, K., Roberts, M. \& Bruce, P. Superstructure control of first-cycle voltage hysteresis in oxygen-redox cathodes. {\em Nature}. \textbf{577}, 502-508 (2020)

\bibitem{maitra2018}Maitra, U., House, R., Somerville, J., Tapia-Ruiz, N., Lozano, J., Guerrini, N., Hao, R., Luo, K., Jin, L., Pérez-Osorio, M. \& Others Oxygen redox chemistry without excess alkali-metal ions in \ce {Na2/3[Mg_{0.28}Mn_{0.72}]O2}. {\em Nat. Chem.}. \textbf{10}, 288 (2018)

\bibitem{wang2018a}Wang, P., Xin, H., Zuo, T., Li, Q., Yang, X., Yin, Y., Gao, X., Yu, X. \& Guo, Y. An Abnormal 3.7 Volt O3-Type Sodium-Ion Battery Cathode. {\em Angew. Chem. Int. Ed.}. \textbf{130}, 8310-8315 (2018)

\bibitem{yabuuchi2014}Yabuuchi, N., Hara, R., Kajiyama, M., Kubota, K., Ishigaki, T., Hoshikawa, A. \& Komaba, S. New O2/P2-type Li-Excess Layered Manganese Oxides as Promising Multi-Functional Electrode Materials for Rechargeable Li/Na Batteries. {\em Adv. Energy Mater.}. \textbf{4}, 1301453 (2014)

\bibitem{cabana2013}Cabana, J., Chernova, N., Xiao, J., Roppolo, M., Aldi, K., Whittingham, M. \& Grey, C. Study of the Transition Metal Ordering in Layered $\rm Na_ {\it x}Ni_{{\it x}/2}Mn_{1- {\it x}/2}O_2$ ($2/3 < {\it x} < 1$) and Consequences of Na/Li Exchange. {\em Inorg. Chem.}. \textbf{52}, 8540-8550 (2013)

\bibitem{song2019}Song, B., Hu, E., Liu, J., Zhang, Y., Yang, X., Nanda, J., Huq, A. \& Page, K. A novel P3-type \ce {Na2/3Mg1/3Mn2/3O2} as high capacity sodium-ion cathode using reversible oxygen redox. {\em J. Mater. Chem. A}. \textbf{7}, 1491-1498 (2019)

\bibitem{hales2011revision}Hales, T., Harrison, J., McLaughlin, S., Nipkow, T., Obua, S. \& Zumkeller, R. A revision of the proof of the Kepler conjecture. {\em The Kepler Conjecture}. pp. 341-376 (2011)

\bibitem{kanyolo2022conformal}Kanyolo, G. \& Masese, T. Advances in Honeycomb Layered Oxides: Part II -- Theoretical advances in the characterisation of honeycomb layered oxides with optimised lattices of cations. {\em ArXiv Preprint ArXiv:2202.10323}. (2022)

\bibitem{kitaev2006anyons}Kitaev, A. Anyons in an exactly solved model and beyond. {\em Annals Of Physics}. \textbf{321}, 2-111 (2006)

\bibitem{kanyolo2020idealised}Kanyolo, G. \& Masese, T. An idealised approach of geometry and topology to the diffusion of cations in honeycomb layered oxide frameworks. {\em Scientific Reports}. \textbf{10}, 1-13 (2020)

\bibitem{kanyolo2022cationic}Kanyolo, G. \& Masese, T. Cationic vacancies as defects in honeycomb lattices with modular symmetries. {\em Scientific Reports}. \textbf{12}, 1-14 (2022)

\bibitem{masese2021mixed}Masese, T., Miyazaki, Y., Rizell, J., Kanyolo, G., Chen, C., Ubukata, H., Kubota, K., Sau, K., Ikeshoji, T., Huang, Z. \& Others Mixed alkali-ion transport and storage in atomic-disordered honeycomb layered $\rm NaKNi_2TeO_6$. {\em Nature Communications}. \textbf{12}, 1-16 (2021)

\bibitem{masese2021topological}Masese, T., Miyazaki, Y., Mbiti Kanyolo, G., Takahashi, T., Ito, M., Senoh, H. \& Saito, T. Topological defects and unique stacking disorders in honeycomb layered oxide $\rm K_2Ni_2TeO_6$ nanomaterials: implications for rechargeable batteries. {\em ACS Applied Nano Materials}. \textbf{4}, 279-287 (2021)

\bibitem{kanyolo2021partition}Kanyolo, G. \& Masese, T. Partition function for quantum gravity in 4 dimensions as a $1/\mathcal{N}$  expansion. preprint: hal-03335930 (2021)

\bibitem{grundish2019electrochemical}Grundish, N., Seymour, I., Henkelman, G. \& Goodenough, J. Electrochemical properties of three $\rm Li_2Ni_2TeO_6$ structural polymorphs. {\em Chemistry Of Materials}. \textbf{31}, 9379-9388 (2019)

\bibitem{kumar2013formation}Kumar, V., Gupta, A. \& Uma, S. Formation of honeycomb ordered monoclinic ${\rm Li_2}M_2{\rm TeO_6}$ ($M = \rm Cu, Ni$) and disordered orthorhombic $\rm Li_2Ni_2TeO_6$ oxides. {\em Dalton Transactions}. \textbf{42}, 14992-14998 (2013)

\bibitem{evstigneeva2011new}Evstigneeva, M., Nalbandyan, V., Petrenko, A., Medvedev, B. \& Kataev, A. A new family of fast sodium ion conductors: ${\rm Na_2}M_2{\rm TeO_6}$ ($M = \rm Ni, Co, Zn, Mg$). {\em Chemistry Of Materials}. \textbf{23}, 1174-1181 (2011)

\bibitem{sankar2014crystal}Sankar, R., Muthuselvam, I., Shu, G., Chen, W., Karna, S., Jayavel, R. \& Chou, F. Crystal growth and magnetic ordering of $\rm Na_2Ni_2TeO_6$ with honeycomb layers and $\rm Na_2Cu_2TeO_6$ with $\rm Cu$ spin dimers. {\em CrystEngComm}. \textbf{16}, 10791-10796 (2014)

\bibitem{berthelot2012studies}Berthelot, R., Schmidt, W., Sleight, A. \& Subramanian, M. Studies on solid solutions based on layered honeycomb-ordered phases P2-${\rm Na_2}M_2{\rm TeO_6}$ ($M = \rm Co, Ni, Zn$). {\em Journal Of Solid State Chemistry}. \textbf{196} pp. 225-231 (2012)

\bibitem{masese2018rechargeable}Masese, T., Yoshii, K., Yamaguchi, Y., Okumura, T., Huang, Z., Kato, M., Kubota, K., Furutani, J., Orikasa, Y., Senoh, H. \& Others Rechargeable potassium-ion batteries with honeycomb-layered tellurates as high voltage cathodes and fast potassium-ion conductors. {\em Nature Communications}. \textbf{9}, 1-12 (2018)

\bibitem{masese2019high}Masese, T., Yoshii, K., Kato, M., Kubota, K., Huang, Z., Senoh, H. \& Shikano, M. A high voltage honeycomb layered cathode framework for rechargeable potassium-ion battery: P2-type $\rm K_{2/3}Ni_{1/3}Co_{1/3}Te_{1/3}O_2$. {\em Chemical Communications}. \textbf{55}, 985-988 (2019)

\bibitem{yoshii2019sulfonylamide}Yoshii, K., Masese, T., Kato, M., Kubota, K., Senoh, H. \& Shikano, M. Sulfonylamide-based ionic liquids for high-voltage potassium-ion batteries with honeycomb layered cathode oxides. {\em ChemElectroChem}. \textbf{6}, 3901-3910 (2019)

\bibitem{tada2022implications}Tada, K., Masese, T. \& Kanyolo, G. Implications of coordination chemistry to cationic interactions in honeycomb layered nickel tellurates. {\em Computational Materials Science}. \textbf{207} pp. 111322 (2022)

\bibitem{kohn1965self}Kohn, W. \& Sham, L. Self-consistent equations including exchange and correlation effects. {\em Physical Review}. \textbf{140}, A1133 (1965)

\bibitem{lobato2021comment}Lobato, A., Salvadó, M. \& Recio, J. Comment on “Uncommon structural and bonding properties in Ag 16 B 4 O 10” by A. Kovalevskiy, C. Yin, J. Nuss, U. Wedig, and M. Jansen, Chem. Sci., 2020, 11, 962. {\em Chemical Science}. \textbf{12}, 13588-13592 (2021)

\bibitem{yin2021reply}Yin, C., Wedig, U. \& Jansen, M. Reply to the ‘Comment on “Uncommon structural and bonding properties in Ag 16 B 4 O 10” by A. Lobato, Miguel Á. Salvadó, and J. Manuel Recio, Chem. Sci., 2021, 12. {\em Chemical Science}. \textbf{12}, 13593-13596 (2021)

\bibitem{vegas2020re}Vegas, A. \& Jenkins, H. A re-interpretation of the structure of the silver borate, Ag16B4O10, in the light of the extended Zintl–Klemm concept. {\em Acta Crystallographica Section B: Structural Science, Crystal Engineering And Materials}. \textbf{76}, 865-874 (2020)

\bibitem{johannes2007formation}Johannes, M., Streltsov, S., Mazin, I. \& Khomskii, D. Formation of an unconventional Ag valence state in $\rm Ag_2NiO_2$. {\em Physical Review B}. \textbf{75}, 180404 (2007)

\bibitem{yoshida2020static}Yoshida, H., Dissanayake, S., Christianson, A., Dela Cruz, C., Cheng, Y., Okamoto, S., Yamaura, K., Isobe, M. \& Matsuda, M. Static and dynamic spin properties in the quantum triangular lattice antiferromagnet $\rm Ag_2CoO_2$. {\em Physical Review B}. \textbf{102}, 024445 (2020)

\bibitem{taniguchi2020butterfly}Taniguchi, H., Watanabe, M., Tokuda, M., Suzuki, S., Imada, E., Ibe, T., Arakawa, T., Yoshida, H., Ishizuka, H., Kobayashi, K. \& Others Butterfly-shaped magnetoresistance in triangular-lattice antiferromagnet $\rm Ag_2CrO_2$. {\em Scientific Reports}. \textbf{10}, 1-7 (2020)

\bibitem{yoshida2011novel}Yoshida, H., Takayama-Muromachi, E. \& Isobe, M. Novel $S = 3/2$ triangular antiferromagnet $\rm Ag_2CrO_2$ with metallic conductivity. {\em Journal Of The Physical Society Of Japan}. \textbf{80}, 123703 (2011)

\bibitem{matsuda2012partially}Matsuda, M., Cruz, C., Yoshida, H., Isobe, M. \& Fishman, R. Partially disordered state and spin-lattice coupling in an $S = 3/2$ triangular lattice antiferromagnet $\rm Ag_2CrO_2$. {\em Physical Review B}. \textbf{85}, 144407 (2012)

\bibitem{yoshida2008unique}Yoshida, H., Ahlert, S., Jansen, M., Okamoto, Y., Yamaura, J. \& Hiroi, Z. Unique phase transition on spin-2 triangular lattice of $\rm Ag_2MnO_2$. {\em Journal Of The Physical Society Of Japan}. \textbf{77}, 074719 (2008)

\bibitem{yoshida2006spin}Yoshida, H., Muraoka, Y., S\"{o}rgel, T., Jansen, M. \& Hiroi, Z. Spin-$1/2$ triangular lattice with orbital degeneracy in a metallic oxide $\rm Ag_2NiO_2$. {\em Physical Review B}. \textbf{73}, 020408 (2006)

\bibitem{zvereva2016orbitally}Zvereva, E., Stratan, M., Ushakov, A., Nalbandyan, V., Shukaev, I., Silhanek, C., Abdel-Hafiez, M., Streltsov, S. \& Vasiliev, A. Orbitally induced hierarchy of exchange interactions in the zigzag antiferromagnetic state of honeycomb silver delafossite $\rm Ag_3Co_2SbO_6$. {\em Dalton Transactions}. \textbf{45}, 7373-7384 (2016)

\bibitem{berthelot2012new}Berthelot, R., Schmidt, W., Muir, S., Eilertsen, J., Etienne, L., Sleight, A. \& Subramanian, M. New layered compounds with honeycomb ordering: $\rm Li_3Ni_2BiO_6$, $\rm {Li_3Ni}M'{\rm BiO_6}$ ($M' =$ Mg, Cu, Zn), and the delafossite $\rm Ag_3Ni_2BiO_6$. {\em Inorganic Chemistry}. \textbf{51}, 5377-5385 (2012)

\bibitem{bette2019crystal}Bette, S., Takayama, T., Duppel, V., Poulain, A., Takagi, H. \& Dinnebier, R. Crystal structure and stacking faults in the layered honeycomb, delafossite-type materials $\rm Ag_3LiIr_2O_6$ and $\rm Ag_3LiRu_2O_6$. {\em Dalton Transactions}. \textbf{48}, 9250-9259 (2019)

\bibitem{bhardwaj2014evidence}Bhardwaj, N., Gupta, A. \& Uma, S. Evidence of cationic mixing and ordering in the honeycomb layer of $\rm Li_4{\it M}SbO_6$ ($M$ (iii) = Cr, Mn, Al, Ga)(S.G. {\it C}2/{\it c}) oxides. {\em Dalton Transactions}. \textbf{43}, 12050-12057 (2014)

\bibitem{beesk1981x}Beesk, W., Jones, P., Rumpel, H., Schwarzmann, E. \& Sheldrick, G. X-ray crystal structure of $\rm Ag_6O_2$. {\em Journal Of The Chemical Society, Chemical Communications}., 664-665 (1981)

\bibitem{ahlert2003ag13oso6}Ahlert, S., Klein, W., Jepsen, O., Gunnarsson, O., Andersen, O. \& Jansen, M. $\rm Ag_{13}OsO_6$: A Silver Oxide with Interconnected Icosahedral $\rm [Ag_{13}]^{4+}$ Clusters and Dispersed $\rm [OsO_6]^{4-}$ Octahedra. {\em Angewandte Chemie}. \textbf{115}, 4458-4461 (2003)

\bibitem{kovalevskiy2020uncommon}Kovalevskiy, A., Yin, C., Nuss, J., Wedig, U. \& Jansen, M. Uncommon structural and bonding properties in $\rm Ag_{16}B_4O_{10}$. {\em Chemical Science}. \textbf{11}, 962-969 (2020)

\bibitem{kohler1985electrical}K\"{o}hler, B., Jansen, M. \& Weppner, W. Electrical properties of some silver-rich ternary oxides. {\em Journal Of Solid State Chemistry}. \textbf{57}, 227-233 (1985)

\bibitem{hull2000structural}Hull, S. \& Keen, D. Structural characterization of the $\beta \rightarrow \alpha$ superionic transition in $\rm Ag_2HgI_4$ and $\rm Cu_2HgI_4$. {\em Journal Of Physics: Condensed Matter}. \textbf{12}, 3751 (2000)

\bibitem{hull2001structural}Hull, S. \& Keen, D. Structural characterization of further high temperature superionic phases of $\rm Ag_2HgI_4$ and $\rm Cu_2HgI_4$. {\em Journal Of Physics: Condensed Matter}. \textbf{13}, 5597 (2001)

\bibitem{hull2002crystal}Hull, S., Keen, D., Sivia, D. \& Berastegui, P. Crystal structures and ionic conductivities of ternary derivatives of the silver and copper monohalides: I. Superionic phases of stoichiometry $\rm MA_4I_5$: $\rm RbAg_4I_5$, $\rm KAg_4I_5$, and $\rm KCu_4I_5$. {\em Journal Of Solid State Chemistry}. \textbf{165}, 363-371 (2002)

\bibitem{hull2004crystal}Hull, S. \& Berastegui, P.Crystal structures and ionic conductivities of ternary derivatives of the silver and copper monohalides—II: ordered phases within the (${\rm Ag}X$)$_x$--($MX$)$_{1-x}$ and (${\rm Cu}X$)$_x$--($MX$)$_{1-x}$ ($M = \rm K, Rb$ and $\rm Cs$; $X = \rm Cl, Br$ and $I$) systems. {\em Journal Of Solid State Chemistry}. \textbf{177}, 3156-3173 (2004)

\bibitem{hull2002structural}Hull, S., Keen, D. \& Berastegui, P. Structural description of the superionic behaviour in the system ${\rm (AgI)}_x$–${\rm (PbI_2)}_{1 - x}$, $2/3 \leq x \leq 4/5$. {\em Solid State Ionics}. \textbf{147}, 97-106 (2002)

\bibitem{nilges2004structure}Nilges, T., Nilges, S., Pfitzner, A., Doert, T. \& B\"{o}ttcher, P. Structure- Property Relations and Diffusion Pathways of the Silver Ion Conductor $\rm Ag_5Te_2Cl$. {\em Chemistry Of Materials}. \textbf{16}, 806-812 (2004)

\bibitem{nilges2005structures}Nilges, T., Dreher, C. \& Hezinger, A. Structures, phase transitions and electrical properties of $\rm Ag_5Te_{2-{\it y}}Se_{\it y}Cl$ ($y = 0 - 0.7$). {\em Solid State Sciences}. \textbf{7}, 79-88 (2005)

\bibitem{matsunaga2004structural}Matsunaga, S. \& Madden, P. Structural and transport properties in the $\rm Ag_3SI$ system: a molecular dynamics study of alpha, beta and molten phases. {\em Journal Of Physics: Condensed Matter}. \textbf{16}, 181 (2004)

\bibitem{hull2005ag+}Hull, S., Berastegui, P. \& Grippa, A. $\rm Ag^+$ diffusion within the rock-salt structured superionic conductor $\rm Ag_4Sn_3S_8$. {\em Journal Of Physics: Condensed Matter}. \textbf{17}, 1067 (2005)

\bibitem{lange2006ag10te4br3}Lange, S. \& Nilges, T. $\rm Ag_{10}Te_4Br_3$: A new silver(I)(poly) chalcogenide halide solid electrolyte. {\em Chemistry Of Materials}. \textbf{18}, 2538-2544 (2006)

\bibitem{lange2007polymorphism}Lange, S., Bawohl, M., Wilmer, D., Meyer, H., Wiemh\"{o}fer, H. \& Nilges, T. Polymorphism, structural frustration, and electrical properties of the mixed conductor $\rm Ag_{10}Te_4Br_3$. {\em Chemistry Of Materials}. \textbf{19}, 1401-1410 (2007)

\bibitem{angenault1989conductivite}Angenault, J., Couturier, J. \& Quarton, M. Conductivite ionique des solutions solides ${\rm Ag}_{1+x}{\rm Zr}_{2-x}M_x{\rm (PO_4)_3}$ avec $M^{III}$ = Sc, Fe. {\em Materials Research Bulletin}. \textbf{24}, 789-794 (1989)

\bibitem{rao2005preparation}Rao, K., Rambabu, G., Raghavender, M., Prasad, G., Kumar, G. \& Vithal, M. Preparation, characterization and impedance study of $\rm AgTa{\it M}P_3O_{12}$ ($M =$ Al, Ga, In, Cr, Fe and Y). {\em Solid State Ionics}. \textbf{176}, 2701-2710 (2005)

\bibitem{daidouh2002structural}Daidouh, A., Durio, C., Pico, C., Veiga, M., Chouaibi, N. \& Ouassini, A. Structural and electrical study of the alluaudites $({\rm Ag}_{1-x}{\rm Na}_x)_2{\rm FeMn_2 (PO_4)_3}$ ($x = 0, 0.5$ and $1$). {\em Solid State Sciences}. \textbf{4}, 541-548 (2002)

\bibitem{daidouh1997structure}Daidouh, A., Veiga, M. \& Pico, C. Structure characterization and ionic conductivity of $\rm Ag_2VP_2O_8$. {\em Journal Of Solid State Chemistry}. \textbf{130}, 28-34 (1997)

\bibitem{fukuoka2003crystal}Fukuoka, H., Matsunaga, H. \& Yamanaka, S. Crystal structure and ionic conductivity of ruthenium diphosphate $A\rm Ru_2(P_2O_7)_2$, $A =$ Li, Na, and Ag, with a tunnel structure. {\em Materials Research Bulletin}. \textbf{38}, 991-1001 (2003)

\bibitem{quarez2009crystal}Quarez, E., Mentre, O., Oumellal, Y. \& Masquelier, C. Crystal structures of new silver ion conductors $\rm Ag_7Fe_3({\it X}_2O_7)_4$ ($\rm {\it X} = P, As$). {\em New Journal Of Chemistry}. \textbf{33}, 998-1005 (2009)

\bibitem{zee2010quantum}Zee, A. Quantum field theory in a nutshell. (Princeton university press,2010)

\bibitem{weinberg1967model}Weinberg, S. A model of leptons. {\em Physical Review Letters}. \textbf{19}, 1264 (1967)

\bibitem{schreyer2002synthesis}Schreyer, M. \& Jansen, M. Synthesis and characterization of $\rm Ag_2NiO_2$ showing an uncommon charge distribution. {\em Angewandte Chemie International Edition}. \textbf{41}, 643-646 (2002)

\bibitem{sorgel2007ag3ni2o4}S\"{o}rgel, T. \& Jansen, M. $\rm Ag_3Ni_2O_4$—A new stage-2 intercalation compound of 2H–$\rm AgNiO_2$ and physical properties of 2H–$\rm AgNiO_2$ above ambient temperature. {\em Journal Of Solid State Chemistry}. \textbf{180}, 8-15 (2007)

\bibitem{gupta2021}Gupta, S. \& Mao, Y. A review on molten salt synthesis of metal oxide nanomaterials: Status, opportunity, and challenge. {\em Progress In Materials Science}. \textbf{117} pp. 100734 (2021)

\bibitem{politaev2010}Politaev, V., Nalbandyan, V., Petrenko, A., Shukaev, I., Volotchaev, V. \& Medvedev, B. Mixed oxides of sodium, antimony (5+) and divalent metals (Ni, Co, Zn or Mg). {\em Journal Of Solid State Chemistry}. \textbf{183}, 684-691 (2010)

\bibitem{pennycook2006}Pennycook, S., Lupini, A., Varela, M., Borisevich, A., Peng, Y., Oxley, M., Benthem, K. \& Chisholm, M. Scanning transmission electron microscopy for nanostructure characterization. {\em Scanning Microscopy For Nanotechnology}. pp. 152-191 (2006)

\bibitem{pennycook1988}Pennycook, S. \& Boatner, L. Chemically sensitive structure-imaging with a scanning transmission electron microscope. {\em Nature}. \textbf{336}, 565-567 (1988)

\bibitem{pennycook2006materials}Pennycook, S., Varela, M., Hetherington, C. \& Kirkland, A. Materials advances through aberration-corrected electron microscopy. {\em MRS Bulletin}. \textbf{31}, 36-43 (2006)

\bibitem{wang2018trapping}Wang, Z., Su, H., Kurmoo, M., Tung, C., Sun, D. \& Zheng, L. Trapping an octahedral \ce {Ag6} kernel in a seven-fold symmetric \ce {Ag56} nanowheel. {\em Nature Communications}. \textbf{9}, 1-8 (2018)

\bibitem{haraguchi2021formation}Haraguchi, N., Okunaga, T., Shimoyama, Y., Ogiwara, N., Kikkawa, S., Yamazoe, S., Inada, M., Tachikawa, T. \& Uchida, S. Formation of Mixed-Valence Luminescent Silver Clusters via Cation-Coupled Electron-Transfer in a Redox-Active Ionic Crystal Based on a Dawson-type Polyoxometalate with Closed Pores. {\em European Journal Of Inorganic Chemistry}. \textbf{2021}, 1531-1535 (2021)

\bibitem{derzsi2021ag}Derzsi, M., Uhliar, M. \& Tokár, K. \ce{Ag6Cl4}: the first silver chloride with rare \ce{Ag6} clusters from an ab initio study. {\em Chemical Communications}. \textbf{57}, 10186-10189 (2021)

\bibitem{jansen1992ag5geo4}Jansen, M. \& Linke, C. \ce{Ag5GeO4}, a new semiconducting oxide. {\em Angewandte Chemie International Edition In English}. \textbf{31}, 653-654 (1992)

\bibitem{jansen1990ag5pb2o6}Jansen, M., Bortz, M. \& Heidebrecht, K. \ce{Ag5Pb2O6}, ein subvalentes Oxid. {\em Journal Of The Less Common Metals}. \textbf{161}, 17-24 (1990)

\bibitem{argay1966redetermination}Argay, G. \& I, N. Redetermination of crystal structure of silver subflouride \ce{Ag2F}. {\em ACTA CHIMICA ACADEMIAE SCIENTARIUM HUNGARICAE}. \textbf{49}, 329 (1966)

\bibitem{bystrom1950crystal}Bystrom, A. \& Evers, L. The crystal structures of \ce {Ag2PbO2} and \ce{Ag5Pb2O6}. {\em Acta Chemica Scandinavica}. \textbf{4}, 613-627 (1950)

\bibitem{molleman2015surface}Molleman, B. \& Hiemstra, T. Surface structure of silver nanoparticles as a model for understanding the oxidative dissolution of silver ions. {\em Langmuir}. \textbf{31}, 13361-13372 (2015)

\bibitem{uchimoto2001changes}Uchimoto, Y., Sawada, H. \& Yao, T. Changes in electronic structure by Li ion deintercalation in $\rm LiNiO_2$ from nickel {\it L}-edge and O {\it K}-edge XANES. {\em Journal Of Power Sources}. \textbf{97} pp. 326-327 (2001)

\bibitem{mccalla2015}McCalla, E., Sougrati, M., Rousse, G., Berg, E., Abakumov, A., Recham, N., Ramesha, K., Sathiya, M., Dominko, R., Van Tendeloo, G. \& Others Understanding the roles of anionic redox and oxygen release during electrochemical cycling of lithium-rich layered \ce {Li4FeSbO6}. {\em Journal Of The American Chemical Society}. \textbf{137}, 4804-4814 (2015)

\bibitem{li2018new}Li, Y., Deng, Z., Peng, J., Gu, J., Chen, E., Yu, Y., Wu, J., Li, X., Luo, J., Huang, Y. \& Others New P2-type honeycomb-layered sodium-ion conductor: $\rm Na_2Mg_2TeO_6$. {\em ACS Applied Materials \& Interfaces}. \textbf{10}, 15760-15766 (2018)

\bibitem{wu2018sodium}Wu, J., Wang, Q. \& Guo, X. Sodium-ion conduction in $\rm Na_2Zn_2TeO_6$ solid electrolytes. {\em Journal Of Power Sources}. \textbf{402} pp. 513-518 (2018)

\bibitem{takada1990}Takada, K., Kanbara, T., Yamamura, Y. \& Kondo, S. Rechargeable solid-state batteries with silver ion conductors. {\em Solid State Ionics}. \textbf{40} pp. 988-992 (1990)

\bibitem{guo2006}Guo, Y., Hu, Y., Lee, J. \& Maier, J. High-performance rechargeable all-solid-state silver battery based on superionic AgI nanoplates. {\em Electrochemistry Communications}. \textbf{8}, 1179-1184 (2006)

\bibitem{inoishi2018}Inoishi, A., Nishio, A., Kitajou, A. \& Okada, S. Single-phase All-solid-state Silver Battery using \ce {Ag_{1.5}Cr_{0.5}Ti_{1.5}(PO4)3} as Anode, Cathode, and Electrolyte. {\em ChemistrySelect}. \textbf{3}, 9965-9968 (2018)

\bibitem{kirshenbaum2016}Kirshenbaum, K., DiLeo, R., Takeuchi, K., Marschilok, A. \& Takeuchi, E. Silver Ion Conducting Electrolytes and Silver Solid-State Batteries. {\em HANDBOOK OF SOLID STATE BATTERIES}. pp. 779-818 (2016)

\bibitem{glukhov2022}Glukhov, A., Belmesov, A., Nechaev, G., Ukshe, A., Reznitskikh, O., Bukun, N., Shmygleva, L. \& Dobrovolsky, Y. Anode material for all-solid-state battery based on solid electrolyte \ce {CsAg4Br_{2.5}I_{2.5}}: Theory and experiment. {\em Materials Science And Engineering: B}. \textbf{278} pp. 115617 (2022)

\bibitem{delaizir2012}Delaizir, G., Manafi, N., Jouan, G., Rozier, P. \& Dollé, M. All-solid-state silver batteries assembled by Spark Plasma Sintering. {\em Solid State Ionics}. \textbf{207} pp. 57-63 (2012)

\bibitem{schmidbaur2015}Schmidbaur, H. \& Schier, A. Argentophilic interactions. {\em Angewandte Chemie International Edition}. \textbf{54}, 746-784 (2015)

\bibitem{jansen1987}Jansen, M. Homoatomic $d^{\rm 10}$-$d^{\rm 10}$ interactions: their effects on structure and chemical and physical properties. {\em Angewandte Chemie International Edition In English}. \textbf{26}, 1098-1110 (1987)

\bibitem{ji2010}Ji, S., Kan, E., Whangbo, M., Kim, J., Qiu, Y., Matsuda, M., Yoshida, H., Hiroi, Z., Green, M., Ziman, T. \& Others Orbital order and partial electronic delocalization in a triangular magnetic metal \ce {Ag2MnO2}. {\em Physical Review B}. \textbf{81}, 094421 (2010)

\bibitem{schwarz2010full}Schwarz, W. The full story of the electron configurations of the transition elements. {\em Journal Of Chemical Education}. \textbf{87}, 444-448 (2010)

\bibitem{blades2017evolution}Blades, W.H., Reber, A.C., Khanna, S.N., L\'{o}pez-Sosa, L., Calaminici, P. \& K\"{o}ster, A.M. Evolution of the Spin Magnetic Moments and Atomic Valence of Vanadium in $\rm VCu_{\it x}^+$, $\rm VAg_{\it x}^+$, and $\rm VAu_{\it x}^+$ Clusters ({\it x} = 3–14). {\em The Journal Of Physical Chemistry A}. \textbf{121}, 2990-2999 (2017)

\bibitem{yang1954conservation}Yang, C. \& Mills, R. Conservation of isotopic spin and isotopic gauge invariance. {\em Physical Review}. \textbf{96}, 191 (1954)

\bibitem{ballhausen1963introduction}Ballhausen, C. \& Weiner, M. Introduction to ligand field theory. {\em Journal Of The Electrochemical Society}. \textbf{110}, 97Cb (1963)

\bibitem{burns1993mineralogical}Burns, R. \& Burns, R. Mineralogical applications of crystal field theory. (Cambridge university press,1993)

\bibitem{huisman1971trigonal}Huisman, R., De Jonge, R., Haas, C. \& Jellinek, F. Trigonal-prismatic coordination in solid compounds of transition metals. {\em Journal Of Solid State Chemistry}. \textbf{3}, 56-66 (1971)

\bibitem{allen2010honeycomb}Allen, M., Tung, V. \& Kaner, R. Honeycomb carbon: a review of graphene. {\em Chemical Reviews}. \textbf{110}, 132-145 (2010)

\bibitem{mecklenburg2011spin}Mecklenburg, M. \& Regan, B. Spin and the honeycomb lattice: lessons from graphene. {\em Physical Review Letters}. \textbf{106}, 116803 (2011)

\bibitem{georgi2017tuning}Georgi, A., Nemes-Incze, P., Carrillo-Bastos, R., Faria, D., Viola Kusminskiy, S., Zhai, D., Schneider, M., Subramaniam, D., Mashoff, T., Freitag, N. \& Others Tuning the pseudospin polarization of graphene by a pseudomagnetic field. {\em Nano Letters}. \textbf{17}, 2240-2245 (2017)

\bibitem{kvashnin2014phase}Kvashnin, A., Chernozatonskii, L., Yakobson, B. \& Sorokin, P. Phase diagram of quasi-two-dimensional carbon, from graphene to diamond. {\em Nano Letters}. \textbf{14}, 676-681 (2014)

\bibitem{grzelak2017metal}Grzelak, A., Gawraczyński, J., Jaroń, T., Kurzydłowski, D., Mazej, Z., Leszczyński, P., Prakapenka, V., Derzsi, M., Struzhkin, V. \& Grochala, W. Metal fluoride nanotubes featuring square-planar building blocks in a high-pressure polymorph of AgF 2. {\em Dalton Transactions}. \textbf{46}, 14742-14745 (2017)

\bibitem{kurzydlowski2021fluorides}Kurzydłowski, D., Derzsi, M., Zurek, E. \& Grochala, W. Fluorides of silver under large compression. {\em Chemistry–A European Journal}. \textbf{27}, 5536-5545 (2021)

\bibitem{ho1990photoelectron}Ho, J., Ervin, K. \& Lineberger, W. Photoelectron spectroscopy of metal cluster anions: Cu-n, Ag-n, and Au-n. {\em The Journal Of Chemical Physics}. \textbf{93}, 6987-7002 (1990)

\bibitem{minamikawa2022electron}Minamikawa, K., Sarugaku, S., Arakawa, M. \& Terasaki, A. Electron counting in cationic and anionic silver clusters doped with a 3d transition-metal atom: endo-vs. exohedral geometry. {\em Physical Chemistry Chemical Physics}. \textbf{24}, 1447-1455 (2022)

\bibitem{dixon1996photoelectron}Dixon-Warren, S., Gunion, R. \& Lineberger, W. Photoelectron spectroscopy of mixed metal cluster anions: NiCu-, NiAg-, NiAg-2, and Ni2Ag-. {\em The Journal Of Chemical Physics}. \textbf{104}, 4902-4910 (1996)

\bibitem{schneider2005unusual}Schneider, H., Boese, A. \& Weber, J. Unusual hydrogen bonding behavior in binary complexes of coinage metal anions with water. {\em The Journal Of Chemical Physics}. \textbf{123}, 084307 (2005)

\bibitem{francesco2012conformal}Francesco, P., Mathieu, P. \& S\'{e}n\'{e}chal, D. Conformal field theory. (Springer Science \& Business Media, 2012)

\bibitem{domb2000phase}Domb, C. Phase transitions and critical phenomena. (Elsevier,2000)

\bibitem{kanyolo2019berry}Kanyolo, G. Berry's Phase and Renormalization of Applied Oscillating Electric Fields by Topological Quasi-Particles. {\em ArXiv Preprint ArXiv:1909.00778}. (2019)

\bibitem{garcia1992dimerization}Garcia-Bach, M., Blaise, P. \& Malrieu, J. Dimerization of polyacetylene treated as a spin-Peierls distortion of the Heisenberg Hamiltonian. {\em Physical Review B}. \textbf{46}, 15645 (1992)

\bibitem{jahn1937stability}Jahn, H. \& Teller, E. Stability of polyatomic molecules in degenerate electronic states-I—Orbital degeneracy. {\em Proceedings Of The Royal Society Of London. Series A-Mathematical And Physical Sciences}. \textbf{161}, 220-235 (1937)

\bibitem{little1964possibility}Little, W. Possibility of synthesizing an organic superconductor. {\em Physical Review}. \textbf{134}, A1416 (1964)

\bibitem{chen2020high}Chen, C., Rizell, J., Kanyolo, G., Masese, T., Sassa, Y., M\aa nsson, M., Kubota, K., Matsumoto, K., Hagiwara, R. \& Xu, Q. High-voltage honeycomb layered oxide positive electrodes for rechargeable sodium batteries. {\em Chemical Communications}. \textbf{56}, 9272-9275 (2020)

\bibitem{masese2021unveiling}Masese, T., Miyazaki, Y., Rizell, J., Kanyolo, G., Takahashi, T., Ito, M., Senoh, H. \& Saito, T. Unveiling structural disorders in honeycomb layered oxide: \ce {Na2Ni2TeO6}. {\em Materialia}. \textbf{15} pp. 101003 (2021)

\bibitem{bera2020temperature}Bera, A. \& Yusuf, S. Temperature-dependent Na-ion conduction and its pathways in the crystal structure of the layered battery material \ce {Na2Ni2TeO6}. {\em The Journal Of Physical Chemistry C}. \textbf{124}, 4421-4429 (2020)

\bibitem{kresse1993ab}Kresse, G. \& Hafner, J. Ab initio molecular dynamics for liquid metals. {\em Physical Review B}. \textbf{47}, 558 (1993)

\bibitem{kresse1994ab}Kresse, G. \& Hafner, J. Ab initio molecular-dynamics simulation of the liquid-metal–amorphous-semiconductor transition in germanium. {\em Physical Review B}. \textbf{49}, 14251 (1994)

\bibitem{kresse1996efficient}Kresse, G. \& Furthmüller, J. Efficient iterative schemes for ab initio total-energy calculations using a plane-wave basis set. {\em Physical Review B}. \textbf{54}, 11169 (1996)

\bibitem{kresse1996efficiency}Kresse, G. \& Furthmüller, J. Efficiency of ab-initio total energy calculations for metals and semiconductors using a plane-wave basis set. {\em Computational Materials Science}. \textbf{6}, 15-50 (1996)

\bibitem{kresse1999ultrasoft}Kresse, G. \& Joubert, D. From ultrasoft pseudopotentials to the projector augmented-wave method. {\em Physical Review B}. \textbf{59}, 1758 (1999)


\end{thebibliography}
\end{document}